\documentclass[a4paper,fleqn,usenatbib,useAMS]{mnras}

\usepackage[british]{babel}             
\usepackage{graphicx}	
\usepackage{amsmath}	
\usepackage{amssymb}	
\usepackage{multicol}        
\usepackage{bm}		
\usepackage{pdflscape}	

\usepackage[usenames]{color}
\usepackage{ulem}

\usepackage[T1]{fontenc}
\usepackage{ae,aecompl}

\usepackage{newtxtext,newtxmath}

\title[The cosmological principle is not in the sky]{The cosmological principle is not in the sky}

\author[C.-G. Park et al.]{
Chan-Gyung Park,$^{1}$\thanks{E-mail: park.chan.gyung@gmail.com}
Hwasu Hyun,$^{2}$
Hyerim Noh$^{3}$\thanks{E-mail: hr@kasi.re.kr}
and Jai-chan Hwang$^{2}$\thanks{E-mail: jchan@knu.ac.kr}
\\
$^{1}$Division of Science Education and Institute of Fusion Science, Chonbuk National University, Jeonju, Korea\\
$^{2}$Department of Astronomy and Atmospheric Sciences, Kyungpook National University, Daegu, Korea\\
$^{3}$Center for Large Telescope, Korea Astronomy and Space Science Institute, Daejeon, Korea
}

\date{Last updated 2015 May 22; in original form 2013 September 5}

\pubyear{2015}

\begin{document}
\label{firstpage}
\pagerange{\pageref{firstpage}--\pageref{lastpage}}
\maketitle

\begin{abstract}
The homogeneity of matter distribution at large scales, known as the cosmological principle, is a central {\it assumption}
in the standard cosmological model.
The case is testable though, thus no longer needs to be a principle.
Here we perform a test for spatial homogeneity using the Sloan Digital Sky Survey Luminous Red Galaxies (LRG) sample by counting galaxies within a specified volume with the radius scale varying up to $300~h^{-1}\textrm{Mpc}$.
We directly confront the large-scale structure data with the definition of spatial homogeneity by {\it comparing} the averages and dispersions of galaxy number counts with allowed ranges of the {\it random} distribution with homogeneity.
The LRG sample shows significantly larger dispersions of number counts than the random catalogues up to $300~h^{-1}\textrm{Mpc}$ scale, and even the average is located far outside the range allowed in the random distribution; the deviations are statistically {\it impossible} to be realized in the random distribution. This implies that the cosmological principle does not hold even at such large scales.
The same analysis of mock galaxies derived from the $N$-body simulation,
however, suggests that the LRG sample is consistent with the current paradigm of cosmology, thus the simulation is also not homogeneous in that scale. We conclude that the cosmological principle is not in the observed sky and nor is demanded to be there by the standard cosmological world model.
This reveals the nature of the cosmological principle adopted in the modern cosmology paradigm, and opens new field of research in theoretical cosmology.
\end{abstract}





\section{Introduction}

The modern physical cosmology is built on a simple geometrical
{\it assumption} about distribution of matter in the large-scale
\citep{Einstein-1917}. The cosmological principle (CP) of modern
cosmology states that spatial distribution of matter is homogeneous
and isotropic in the large-scale. Being a statement on physical
state of matter, it is testable using the observed redshift and
the angular location in the sky of luminous galaxies;
the distance is not directly available and in cosmology we have
to consider that different distances correspond to different temporal
epochs in the history of the universe. The physical state of
distribution of course cannot be exactly homogeneous and isotropic
which is merely a mathematical idealization.
In practice however one can {\it test} the assumption by comparing
the distribution with the random one. As we know that the small
scale distributions of celestial objects are apparently far from
homogeneous or isotropic, we can anticipate that if the CP is true
there might appear a homogeneity-scale (HS) above which the
distribution is statistically indistinguishable from the random one.
Our aim in this paper is to find out the HS from observation.

Such studies based on galaxy count are available in the literature and there are several claims about the HS reached at scales of $70$--$100~h^{-1}\textrm{Mpc}$
(\citealt{hogg,scrimgeour,ntelis,yadav,laurent,bagla,
sarkar}; see \citealt{pandey-150,yadav-260} for somewhat larger HS);
$h$ is the Hubble parameter normalized to $100~\textrm{km}\textrm{s}^{-1}\textrm{Mpc}^{-1}$; for contrary views, see \cite{labini-3,labini-4,labini-5}.
We find, however, most of the methods used to reach the
conclusion are not about direct test one can na\"ively think of,
i.e, comparing the distribution (like the average and dispersion) in the galaxy counts with the one in the {\it random}
realization. Our result based on such a method shows that even far
beyond the claimed HS the observed distribution is {\it not} homogeneous
even at $300~h^{-1}\textrm{Mpc}$ scales!
This is true not only for dispersions but also for average as well.
Therefore, our conclusion is that the observation we have used does
{\it not} show any HS.

Our conclusion does not necessarily imply that the current cosmological paradigm is seriously challenged.
Although the modern cosmology is fundamentally based on the CP, it is a {\it theoretical} assumption on a {\it fictitious} background.
In a realistic case the background model should be added by small-amplitude (about $2 \times 10^{-5}$ dimensionless level in the early epoch) perturbations in all scales, and the background can be achieved by spatial averaging.
Due to the gravitational instability the fluctuations are amplified in time especially from the small scales.
Thus, as long as the consequent theoretically predicted fluctuations are consistent with the observation the cosmological paradigm based on CP in the background universe (in the early era) is safe independently of the strong statement on the actual existence of CP in the observed sky; without the HS the CP in that space (even in a background) lose its meaning.

Although the issue on whether or not the colossal structures discovered in the large-scale, like the Sloan Digital Sky Survey (SDSS) Great Wall boasting
$300~h^{-1}\textrm{Mpc}$ in linear dimension, is consistent with the cosmological simulation is currently under debate \citep{park}, our additional test in this work, now comparing observation with the simulation, reveals that the observed fluctuations are {\it consistent} with the simulated ones at radius scales up to $300~h^{-1}\textrm{Mpc}$.
Thus, we may conclude that although we do not have the HS in the observed galaxy distribution, the modern cosmology theoretically based on the CP is consistent with the observed large-scale galaxy distribution.

In this paper, we perform a direct test for spatial homogeneity of large-scale structure using the recent galaxy redshift survey data by counting galaxies within a specified volume with varying size scale.
Unlike the previous studies based on the {\it average trend} of galaxy counts over different scales, our analysis {\it compares} the average and dispersion of galaxy number counts {\it with} the expected range allowed by the random distribution with homogeneity.

The outline of this paper is as follows.
In Section 2, we describe the large-scale structure data together with random and mock catalogues used in our analysis.
In Sections 3 and 4, we count galaxies within a sphere and within redshift ranges, respectively,
and compare the results with those from the random and mock data sets.
We discuss our results in Section 5 and present the conclusion in Section 6.
Appendix includes technical details for the data and the analysis methods.

\begin{figure}
\centering
\includegraphics[width=85mm]{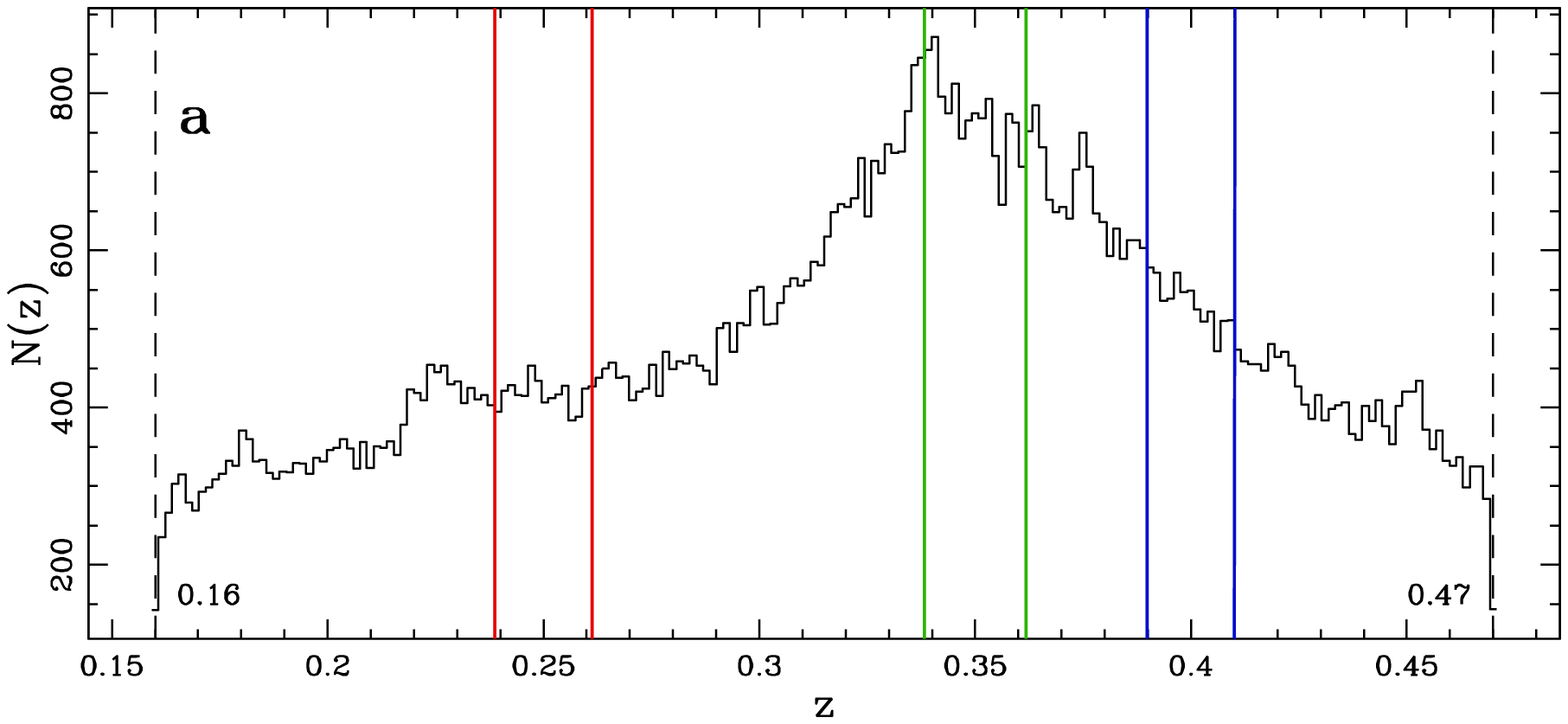}
\includegraphics[width=80mm]{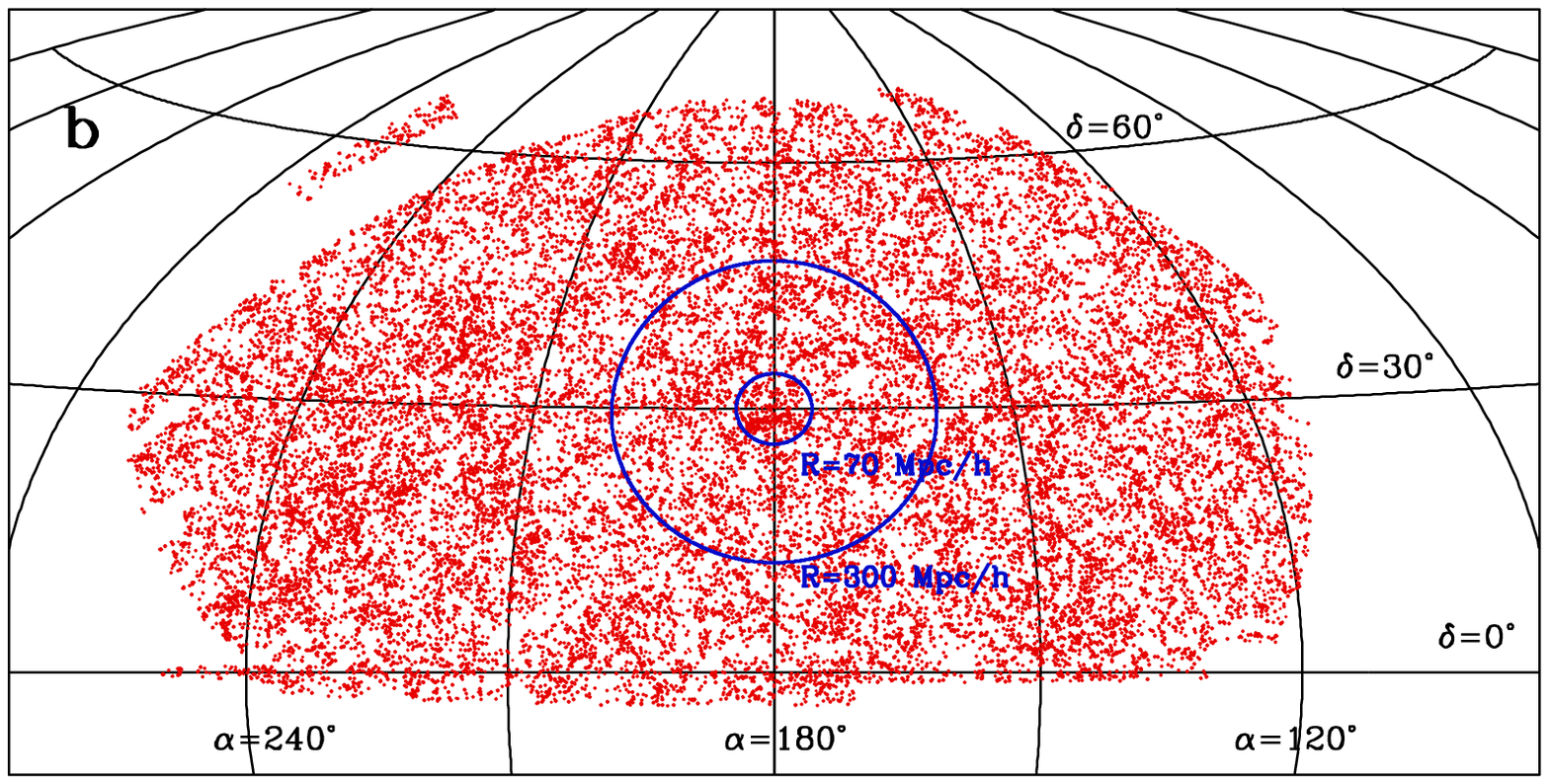}
\caption{($a$) Redshift and ($b$) angular distributions of the SDSS LRG.
         In panel $a$, vertical lines denote three redshift slices of $50~h^{-1}\textrm{Mpc}$ thickness
         centred at $z_c=0.25$, $0.35$, and $0.40$. In panel $b$, angular distribution is shown for galaxies within
         a slice with thickness of $140~h^{-1}\textrm{Mpc}$ centred at $z_c=0.35$
         in the Hammer-Aitoff equal-area projection with equatorial coordinates.
         Circles with the claimed HS ($70~h^{-1}\textrm{Mpc})$ and
         $300~h^{-1}\textrm{Mpc}$ as radius
         are shown for a comparison.
         }
\label{fig:lrg}
\end{figure}

%
%
%
\section{Data}

We use the SDSS Data Release 7 Luminous Red Galaxies (LRG) sample
\citep{sdss-lrg,eisenstein} to explore the homogeneity of our local universe.
Although the SDSS LRG sample is smaller in galaxy number and survey volume than the recent galaxy surveys, it is still useful for homogeneity test due to two reasons. First, we can compare our result with the previous ones obtained with the same sample \citep{hogg}. Second, since there is a claimed HS at $70~h^{-1}\textrm{Mpc}$ scales that is far smaller than the LRG survey size, the larger sample of galaxies is not needed to draw a counter result.
The LRG sample contains galaxies that are believed to be a good
tracers of massive halos, and quasi-volume-limited up to redshift
$z\simeq 0.36$, thereafter flux-limited up to $z\simeq 0.47$.
We use the LRG sample provided in \cite{kazin},
which includes 105,831 galaxies over redshift range of $0.16 < z < 0.47$
with effective volume of $1.6~h^{-3}\textrm{Gpc}^3$
(Fig.\ \ref{fig:lrg}; see Appendix).

In order to decide whether or not the LRG are homogeneously
distributed at a given scale, we compare the galaxy counts with
those from homogeneous distribution.
For this purpose, we generate 1,000 random catalogues, each containing
the Poisson-distributed data points with the same number as in the LRG sample,
by considering LRG redshift distribution and
angular selection function (Fig.\ \ref{fig:angsel}) as the probability functions.
We analyse these catalogues to set the criterion for the spatial homogeneity.

We also make the mock catalogues that mimic the LRG sample.
Using the all-sky lightcone halo catalogues made from 
$N$-body simulation \citep{kim}, we generate 1,296 LRG mock data sets,
with individual survey area uniformly separated but significantly overlapped on the sky.
Within the survey area, massive halos with the same number of LRG have
been extracted in decreasing order of halo mass by considering
the LRG redshift distribution and
angular selection function (see Appendix for details).
We analyse these mock data sets to check whether or not the observation
is consistent with the current paradigm of cosmology.

\begin{figure*}
\includegraphics[width=88mm]{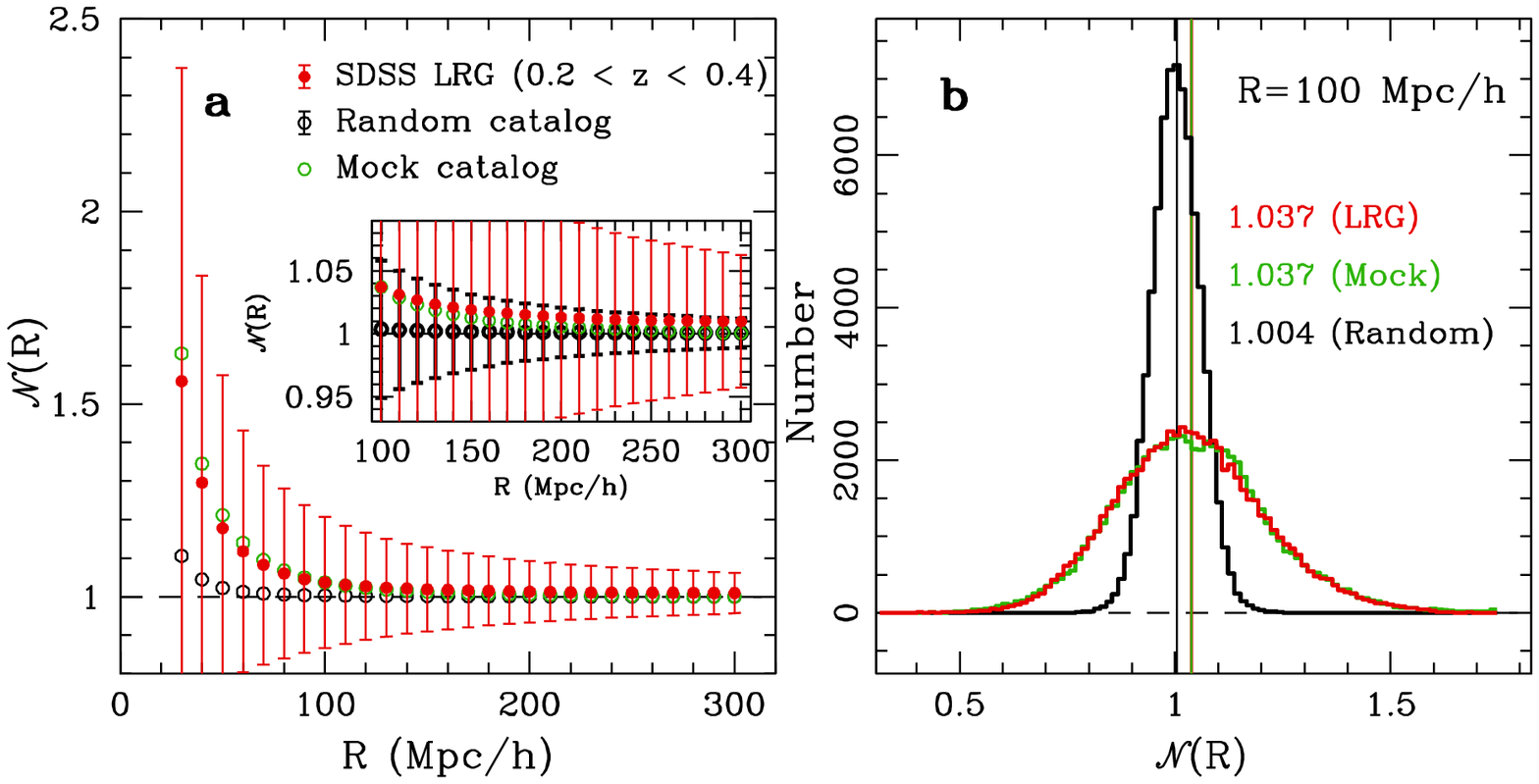}
\includegraphics[width=88mm]{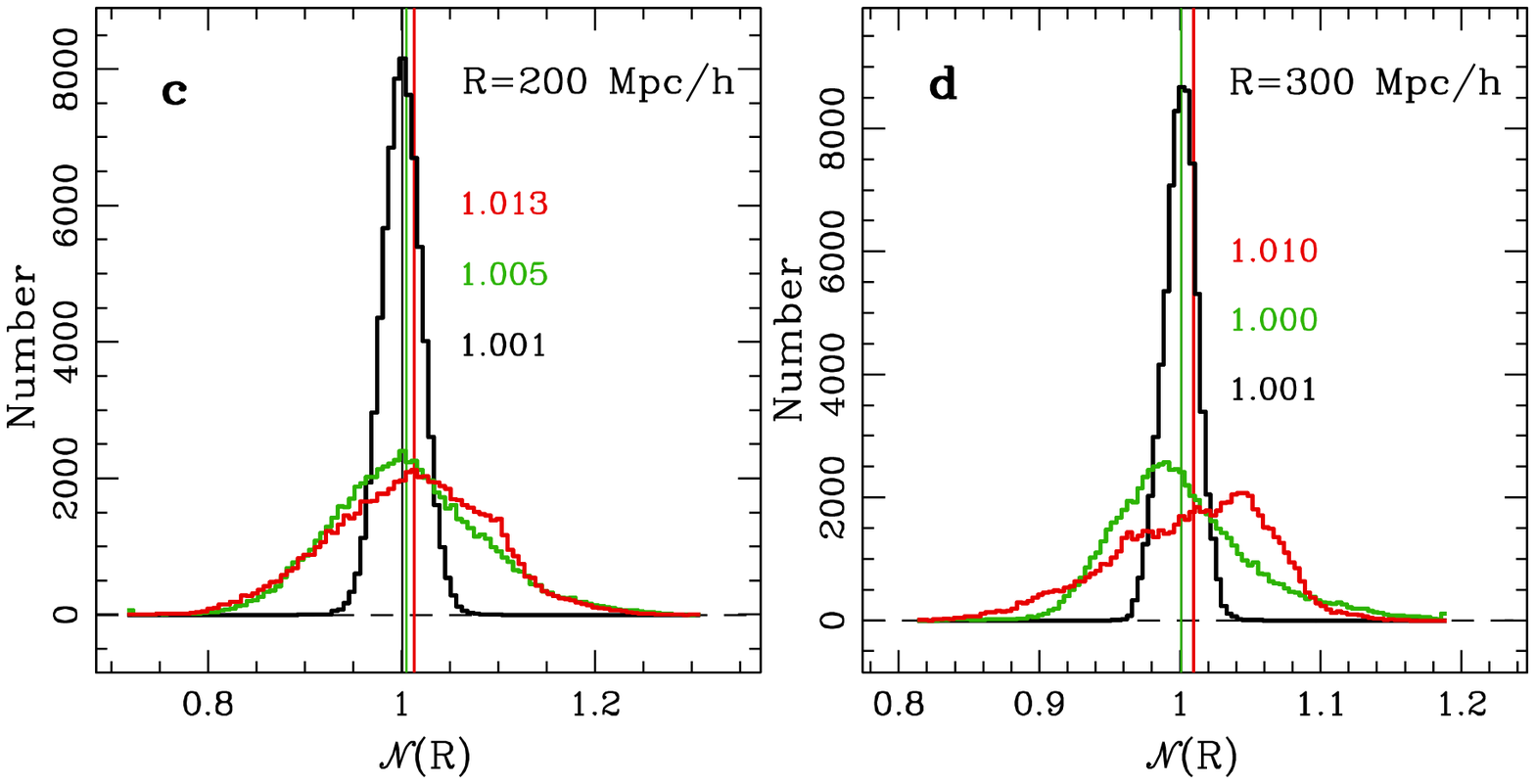}
\includegraphics[width=88mm]{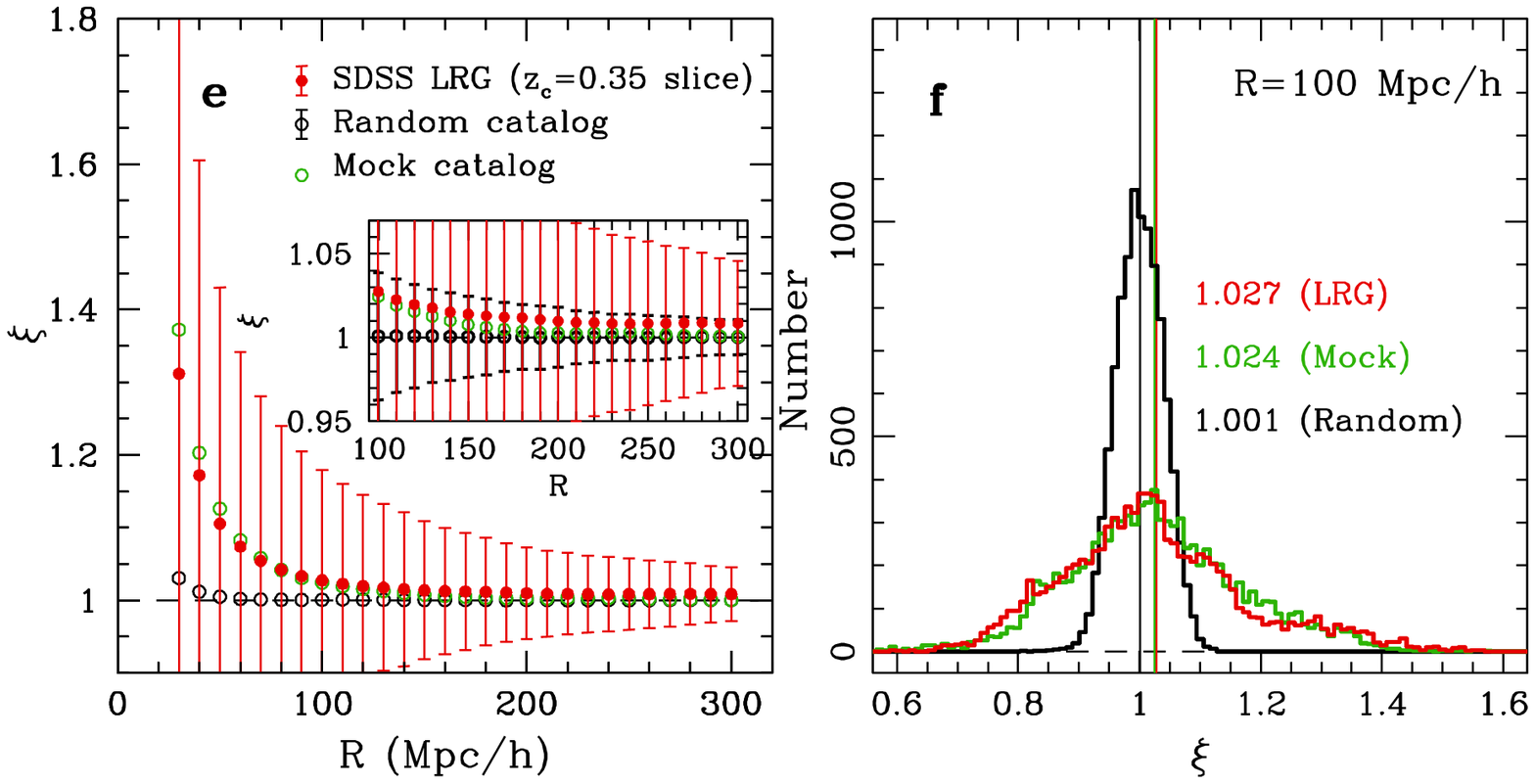}
\includegraphics[width=88mm]{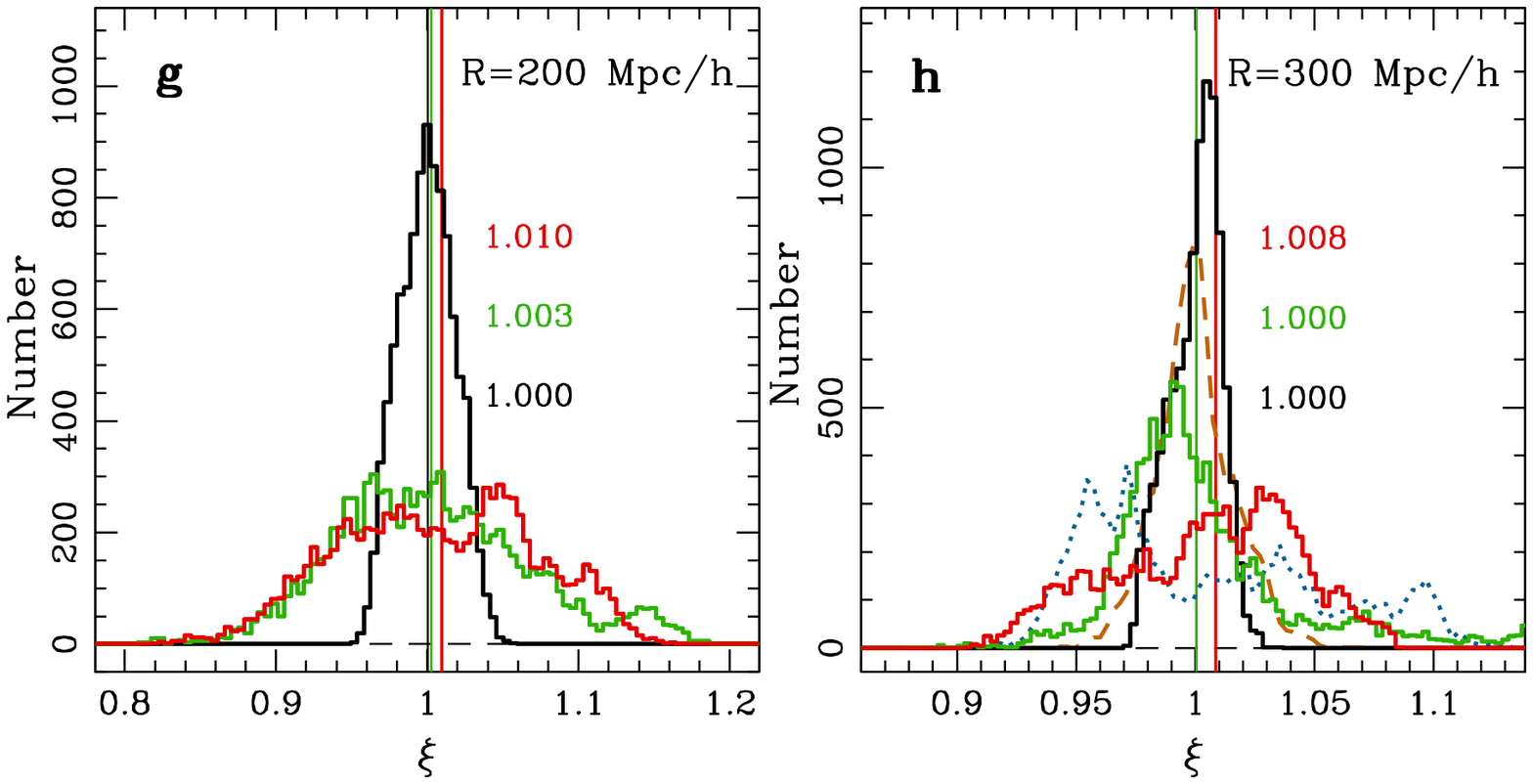}
\caption{Results of counting galaxies over varying size scales.
         ($a$) The weighted average of the scaled counts-in-sphere
         $\mathcal{N}(R)$ versus $R=30$--$300~h^{-1}\textrm{Mpc}$,
         estimated from the LRG at $0.2 < z < 0.4$ (red dots).
         Black and green circles indicate results for one of the random and mock
         catalogues, respectively.
         Error bars indicate the standard deviations
         of $\mathcal{N}(R)$'s, and those for the LRG and a random catalogue are compared in the small panel.
         ($b$--$d$) Histograms of individual $\mathcal{N}$'s
         at $R=100$, $200$, and $300~h^{-1}\textrm{Mpc}$, with the same colour code.
         Vertical lines with numbers (from top to bottom) indicate the average of
         $\mathcal{N}$'s from the LRG, mock and random catalogues.
         ($e$--$h$) The same as panels $a$--$d$ but for $\xi$ measurements
         for the LRG within the $z_c=0.35$ slice.
         For $R=300~h^{-1}\textrm{Mpc}$ in panel $h$,
         the blue dotted and brown dashed curves indicate the results for the mock
         catalogues having the maximum and minimum dispersions in the $\xi$-distribution, respectively
         (see Figs.\ \ref{fig:xi-maps}$c$--$d$ and \ref{fig:xi_mean_std}$f$).}
 \label{fig:count}
\end{figure*}

%
%
%
\section{Counting galaxies within a sphere}

To test the homogeneity of large-scale structure, first we apply the
count-in-sphere method in the similar way as in \cite{hogg}.
As a measure of homogeneity, we calculate the scaled count-in-sphere
$\mathcal{N}(R)$ defined as the number of galaxies within a sphere of
comoving radius $R$ centred at each galaxy divided by the number
expected in the homogeneous distribution.
The latter is estimated from a random point distribution which contains
100 times larger number of data points than the single data set.


Figure \ref{fig:count}$a$--$d$ show the weighted average of the
scaled counts for comoving radius up to $300~h^{-1}\textrm{Mpc}$
and histograms of individual $\mathcal{N}$'s for three chosen radii.
Here we use all the galaxies at $0.2 < z < 0.4$ in the LRG sample
as the centre of sphere. For galaxies around the edge of the survey region,
we assign the smaller weight in estimating the weighted average and standard deviation
of the scaled $\mathcal{N}$'s, where the volume-completeness is used as the weight.
The volume-completeness for each measured $\mathcal{N}$ is defined as
a fraction of the (partial) volume of the sphere contained within the survey region to
that of a complete sphere, centred at the location of a galaxy (see Appendix for estimating the partial volume of a sphere).
For comparison, we have also analysed one random and one mock catalogues in the same way as the LRG catalogue has been analysed. 

For the LRG sample, the averaged $\mathcal{N}$ goes over into unity with a flat slope
at scales larger than around $70~h^{-1}\textrm{Mpc}$, approaching to
$\mathcal{N}=1$ within 1\% at $R=300~h^{-1}\textrm{Mpc}$,
which is very similar to the results shown in \cite{hogg}
and used to imply the existence of claimed HS.
However, individual scaled counts show significant dispersions deviating
from the average. The scatter decreases as the radius increases
with standard deviations of $0.17$ ($0.05$), $0.08$ ($0.02$) and
$0.05$ ($0.01$) for the LRG (one random) catalogue at $R=100$, $200$ and
$300~h^{-1}\textrm{Mpc}$, respectively.
Thus, the LRG sample has 5 times larger dispersions than the random catalogue even at $300~h^{-1}\textrm{Mpc}$.


We emphasize that {\it dispersions} are more important measure of the homogeneity than the approaching of {\it average} numbers to the homogeneous ones; the latter is only one of the necessary conditions for homogeneity.
Later we will show that even the average of the LRG (while consistent with mock data) show statistically significant deviation from the random one.

\begin{figure}
\includegraphics[width=85mm]{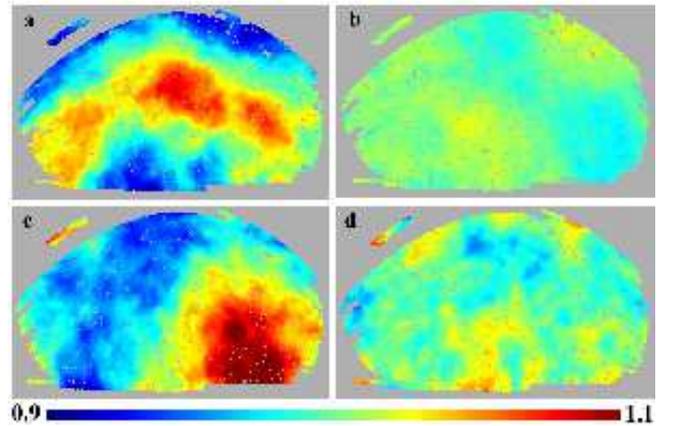}
\caption{Angular distributions of $\xi$.
         Shown are $\xi$ maps for the $z_c=0.35$ slice
         at $R=300~h^{-1}\textrm{Mpc}$ scale estimated
         from ($a$) the LRG, ($b$) one random,
         and mock catalogues with ($c$) the maximum and
         ($d$) minimum dispersions in $\xi$ measurements;
         locations of $b$--$d$ in a statistics plot are
         indicated as grey symbols in Fig.\ \ref{fig:xi_mean_std}$f$.
         The corresponding angular distributions
         of data points are shown in Fig.\ \ref{fig:lrg_R300}.
         }
\label{fig:xi-maps}
\end{figure}

\begin{figure*}
\centering
\includegraphics[width=43mm]{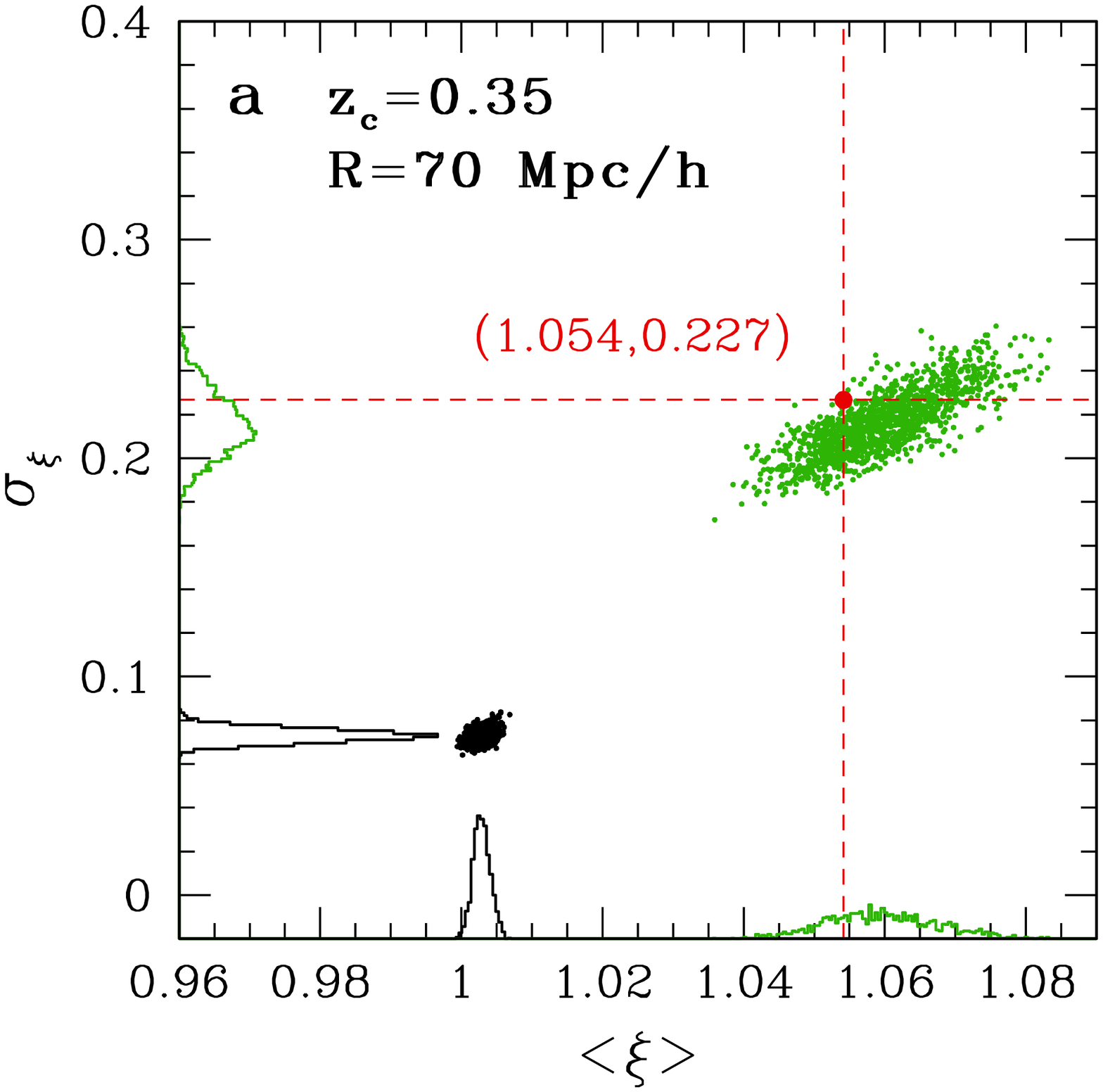}
\includegraphics[width=43mm]{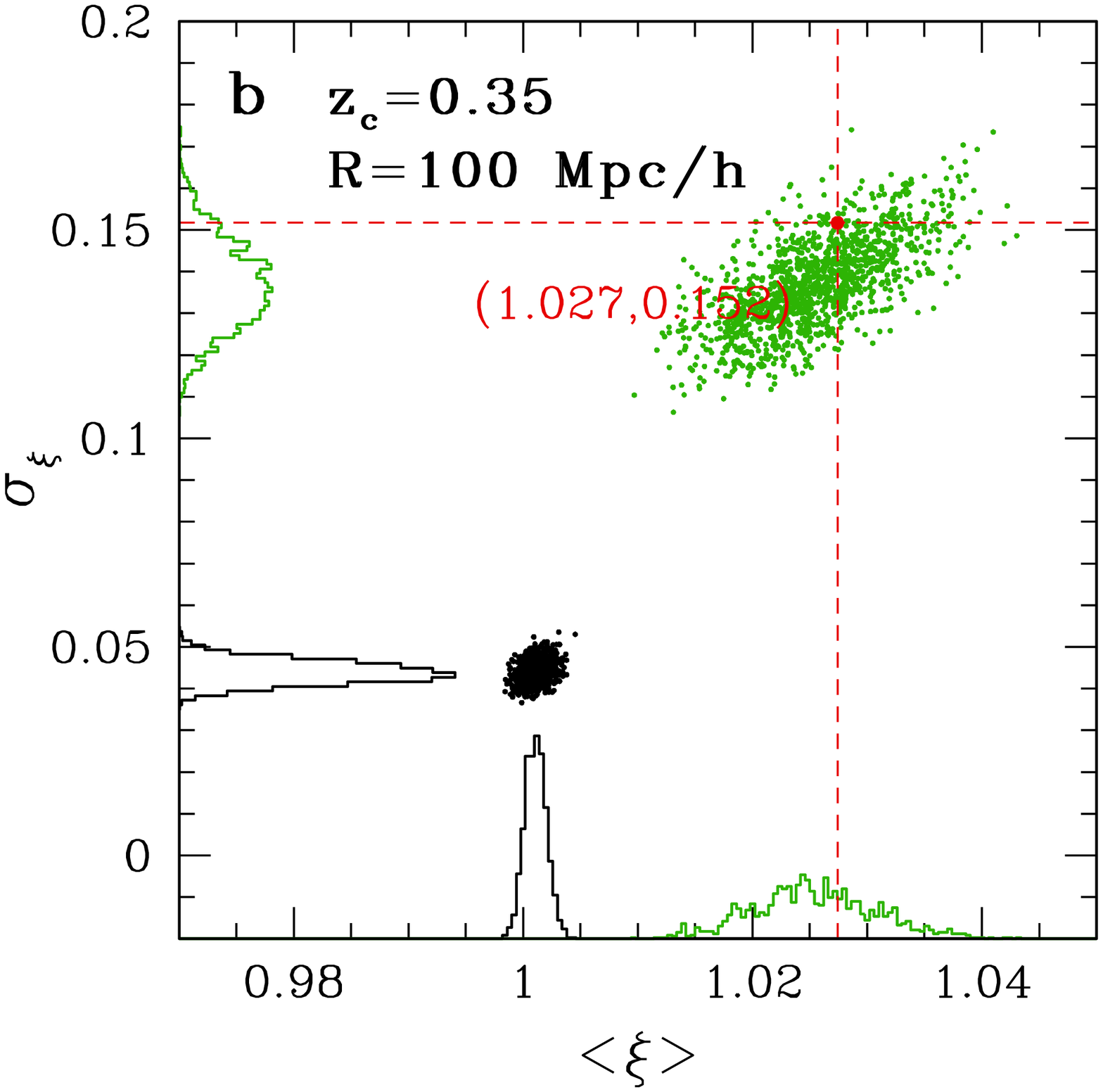}
\includegraphics[width=43mm]{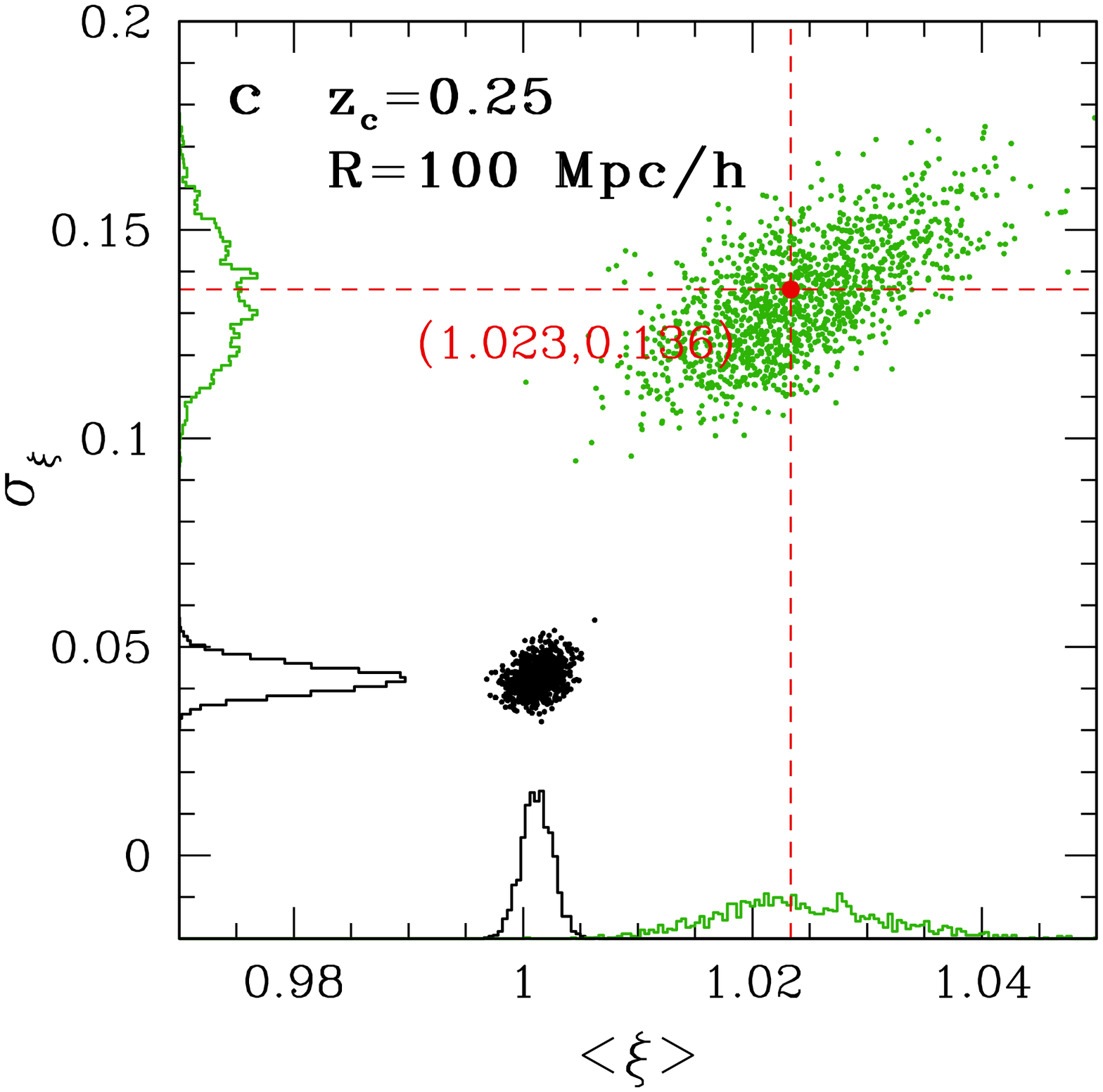}
\includegraphics[width=43mm]{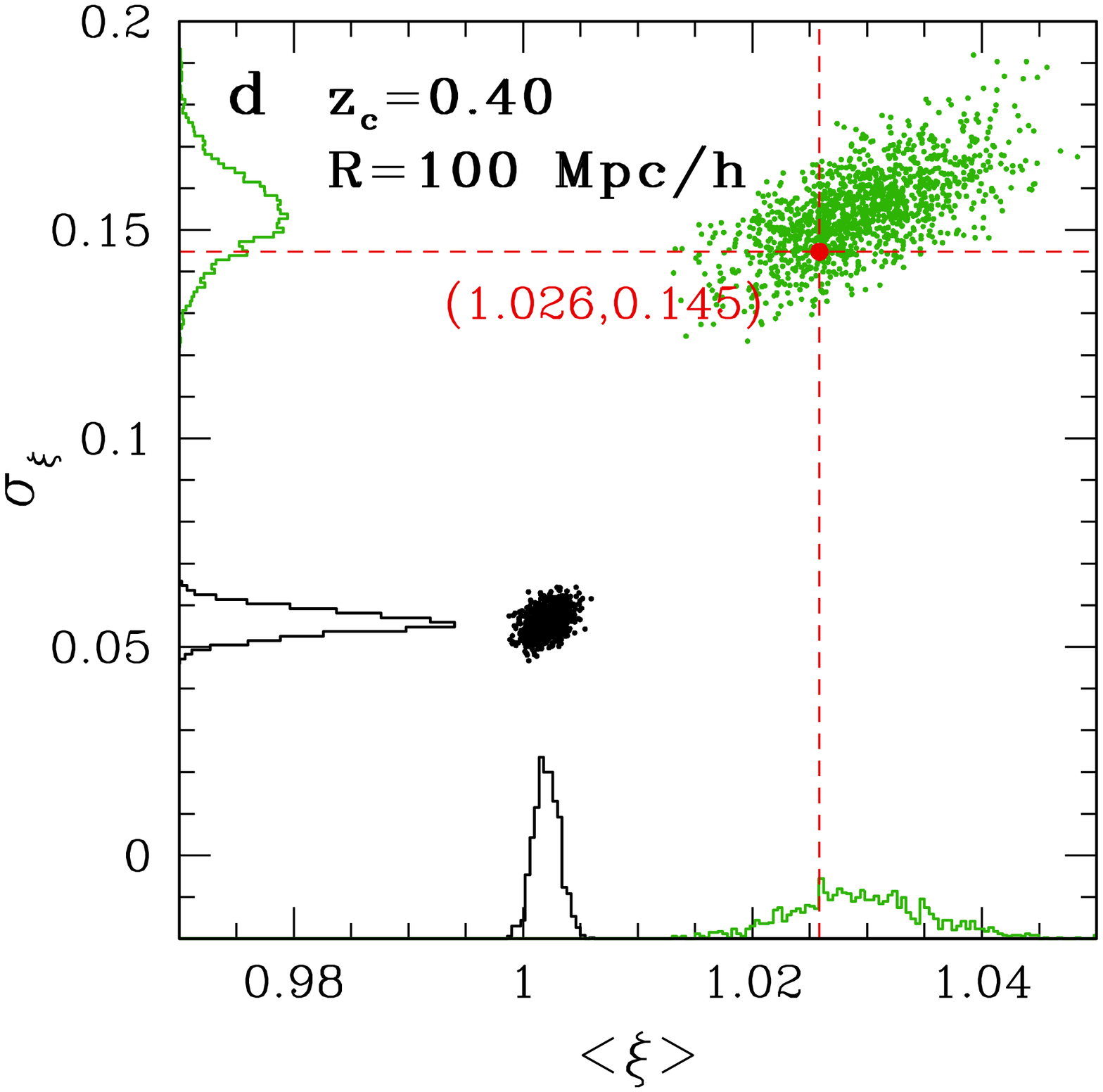}
\includegraphics[width=43mm]{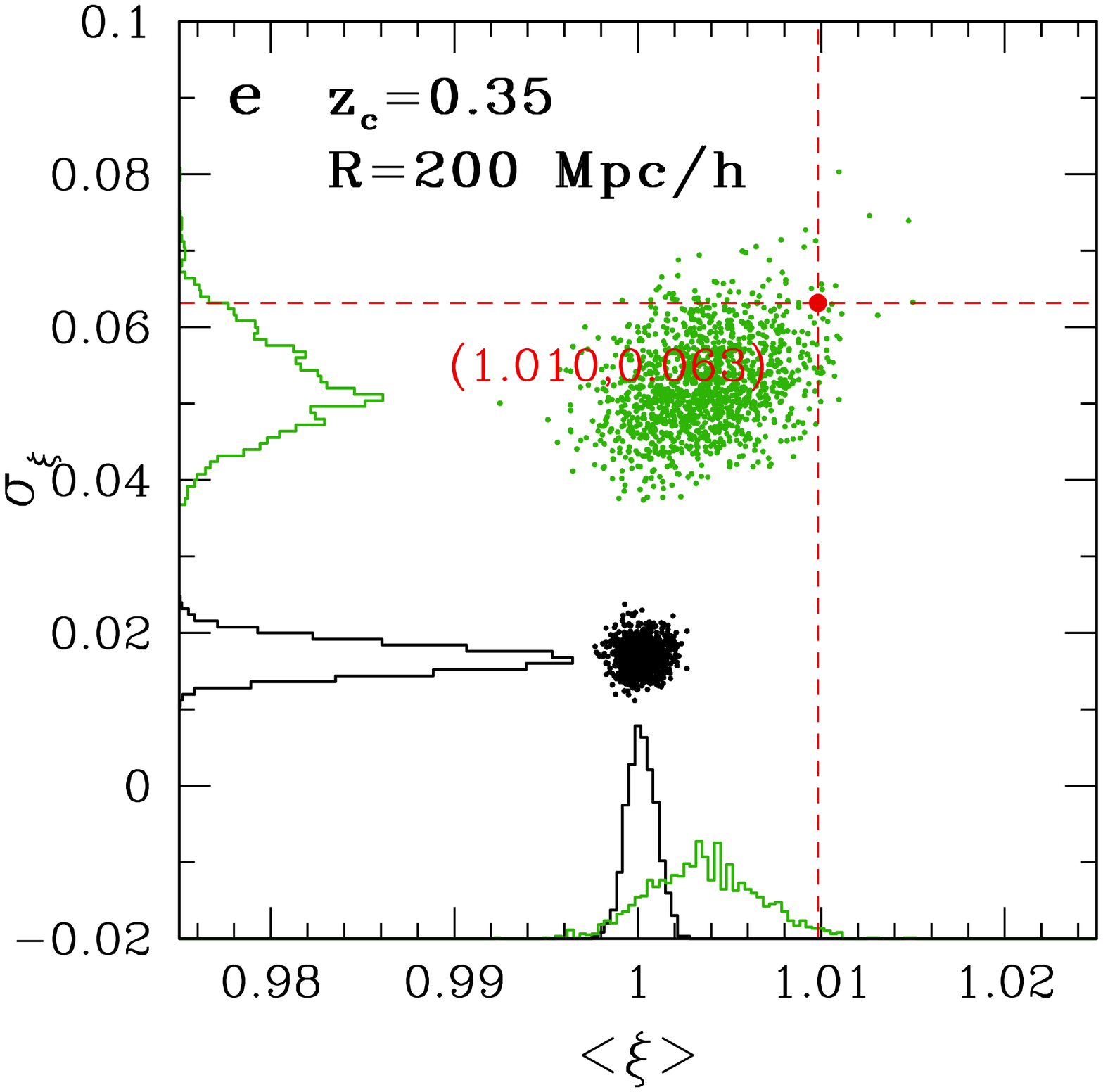}
\includegraphics[width=43mm]{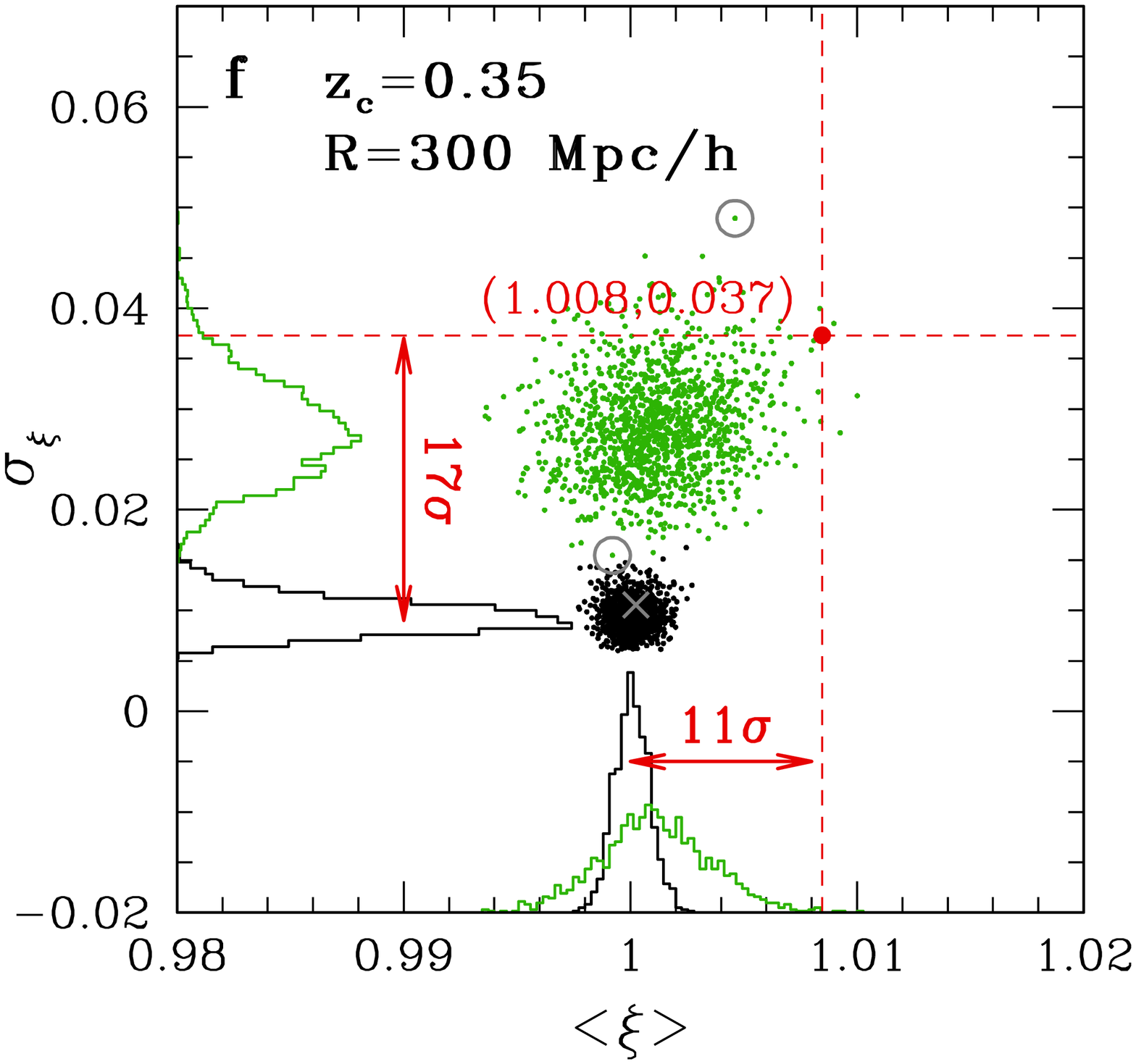}
\includegraphics[width=43mm]{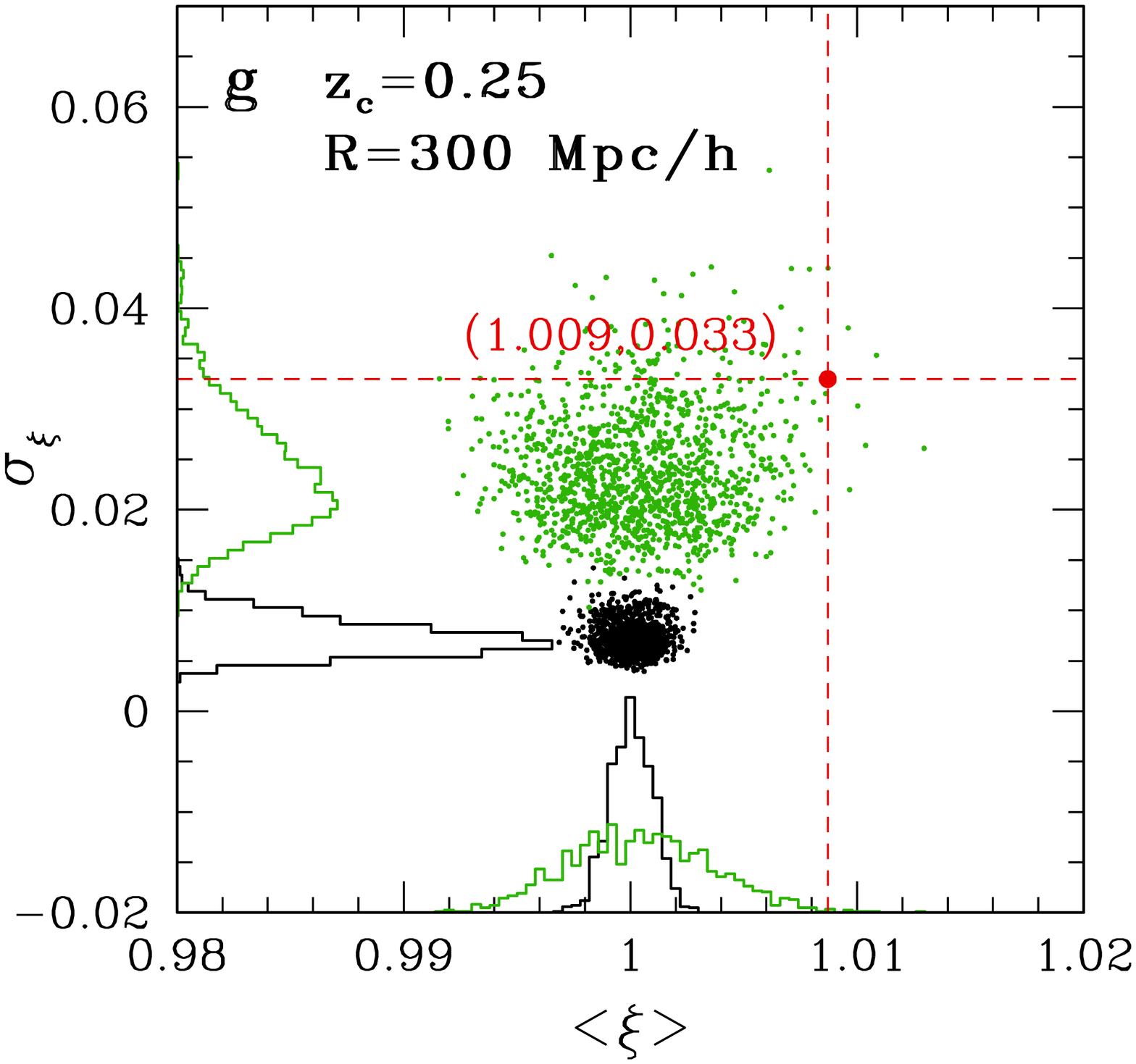}
\includegraphics[width=43mm]{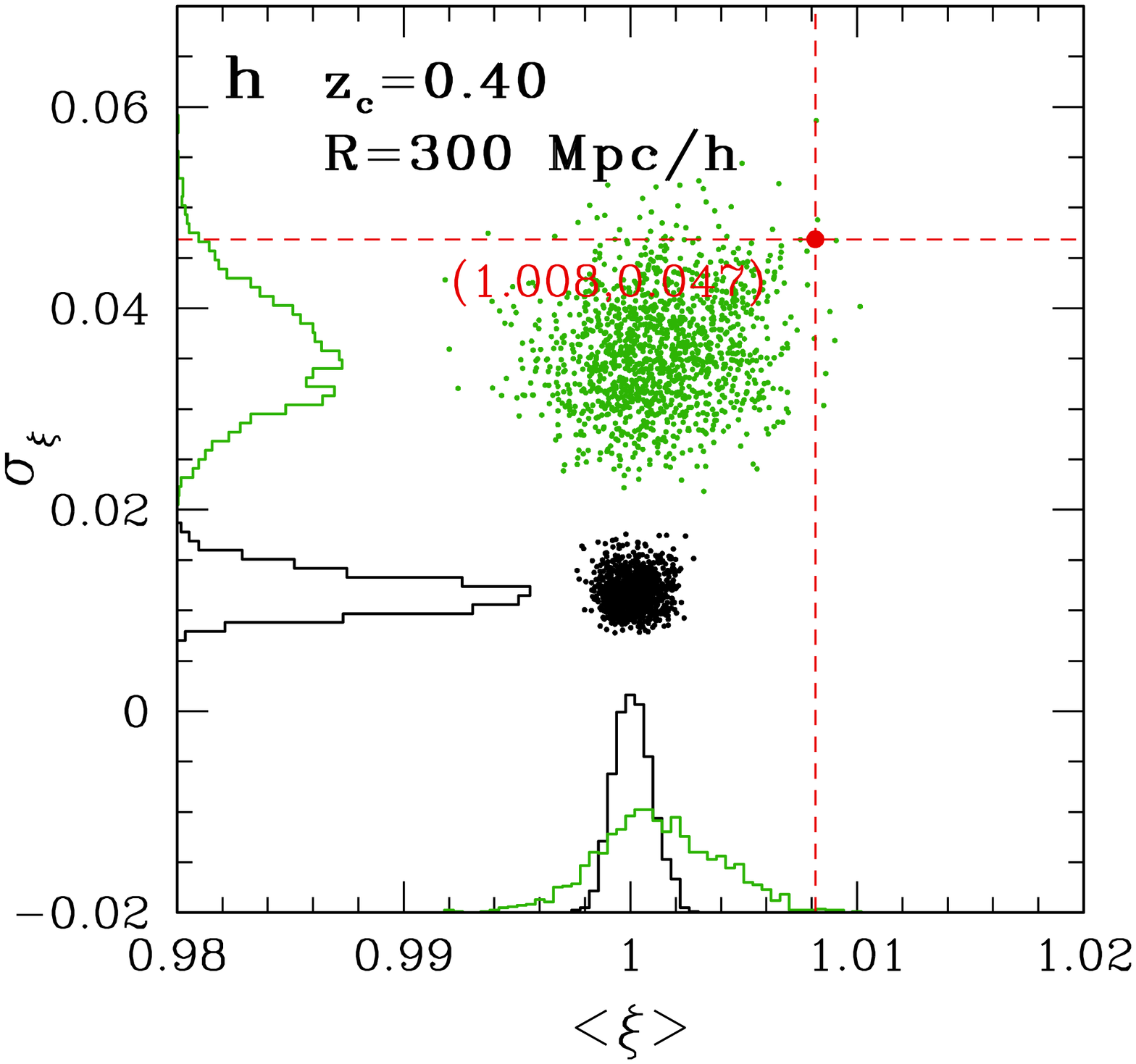}
\caption{Plots of $\left<\xi\right>$ versus $\sigma_\xi$.
         Results for the $z_c=0.35$ slice at $R=70$, $100$, $200$,
         $300~h^{-1}\textrm{Mpc}$ are shown in panels $a$, $b$, $e$, $f$, respectively,
         estimated from the LRG (red dots with dashed lines and numbers),
         1,000 random (black), and 1,296 mock catalogues (green dots),
         with the corresponding histograms shown on the axes.
         In panel $f$, grey symbols indicate the random catalogue and two
         mock catalogues (with the maximum and minimum $\sigma_\xi$)
         that were chosen to present
         Fig.\ \ref{fig:xi-maps}$b$--$d$.
         Panels $c$--$d$ and $g$--$h$ present the results for
         $z_c=0.25$ and $0.40$ slices at $R=100$ and $300~h^{-1}\textrm{Mpc}$,
         respectively.
         }
\label{fig:xi_mean_std}
\end{figure*}


%
%
%
\section{Counting galaxies within redshift ranges}

The count-in-sphere method needs the random point distribution to correct
for the bias due to the survey incompleteness in radial (thus time) direction.
Here we apply a count-in-redshift-range method which avoids
such a bias without the need to use the random distribution.
In order to save computation time we consider a truncated cone instead of a sphere.
We select LRG within a slice of $50~h^{-1}\textrm{Mpc}$ thickness
at the central redshift $z_c$.
In our analysis three slices are chosen with $z_c=0.25$, $0.35$, and $0.40$
(Fig.\ \ref{fig:lrg}$a$).
For each galaxy within the thin slice, we place a sphere of radius $R$
at $z_c$ but with the galaxy's angular position,
and define a truncated cone that is circumscribed about the sphere (see Fig.\ \ref{fig:geom_trc} in the Appendix).
The upper and base sides of the truncated cone set
the minimum and maximum redshifts ($z_1$ and $z_2$) which determine
the slice of $z_2-z_1$ (or $2R$) thickness.
Then, we count galaxies within the truncated cone and calculate
a new measure of homogeneity $\xi$ defined as the number density of galaxies
within a truncated cone divided by that within the whole slice:
\begin{equation}
   \xi (R) = \frac{N_{\textrm{trc}}/V_{\textrm{trc}}}{N_{\textrm{slice}}/V_{\textrm{slice}}}
\end{equation}
where $N_{\textrm{trc}}$ is the number of galaxies within the truncated cone
with volume $V_{\textrm{trc}}$ while $N_{\textrm{slice}}$ is the number of galaxies
within the slice at redshift range $z_1 < z < z_2$ covering the survey's whole angular area with volume $V_{\textrm{slice}}$.

Although the $\xi$-statistic is independent of the bias due to the radial selection function (or the redshift distribution), the random catalogues are still needed for homogeneity test of the LRG sample because the homogeneity expectation can be quantified only by analysing the random catalogues.
For homogeneous distribution, we expect the individual $\xi$ (not the averaged quantity) to approach unity within the precision of homogeneity expectation at scales larger than HS.

As in the case of the count-in-sphere, the volume of the truncated cone centred at a galaxy located
around the edge of the survey or near the masked area of bright stars and bad fields is not complete but partial due to the survey boundary and mask. In this work, we accurately estimate the partial volume of the truncated cone within the survey region by integrating the volume elements over the pixelised angular area enclosed by the partial truncated cone. We define the volume-completeness ($P$) as the partial volume of the truncated cone divided by the complete volume of the same cone (see Appendix for estimating the partial and complete volume of the truncated cone). 

Figure \ref{fig:count}$e$--$h$ show the weighted average of $\xi$'s
versus $R$ for $z_c=0.35$ slice galaxies,
together with histograms of individual $\xi$'s for three chosen radii.
The volume-completeness is used as a weight for each $\xi$ measurement in estimating the average and standard deviation. For comparison, we also present the results for one random and mock catalogues that were analysed in the same way as the LRG catalogue.
As in the case of count-in-sphere, the LRG histograms from the
count-in-redshift-range analysis confirm that the averaged $\xi$ seems
to approach homogeneity with 1\% accuracy at $300~h^{-1}\textrm{Mpc}$ scale.
However, individual $\xi$'s show significant dispersions far deviating
from the range allowed by the random catalogues with homogeneous distribution.
The unstable behaviours of dispersion for LRG and mock data
at $300~h^{-1}\textrm{Mpc}$ scale also suggest that the homogeneity has not
been reached yet at such a scale (Fig.\ \ref{fig:count}$h$).
This is visually demonstrated in Fig.\ \ref{fig:xi-maps}
where $\xi$ has been estimated on pixelised positions within the survey area
for the LRG, one random, two mock catalogues with the maximum and minimum
dispersions in $\xi$ measurements.


%
%
\section{Discussion}

In this paper, we directly confront the large-scale structure data
with the definition of spatial homogeneity by considering that beyond the HS there
would be no variation in the galaxy counts within the scatter expected
in the Poisson distribution.
The two galaxy-counting methods we adopted have a limitation
that $\mathcal{N}$ (or $\xi$) goes over into unity on the survey-sized
scale regardless of whether or not homogeneity has been reached
\citep{scrimgeour}.
However, for a distribution with HS smaller than the survey size
the $\mathcal{N}$ and $\xi$ should approach unity at all scales
beyond HS within the precision allowed by the random catalogues.
Our analyses demonstrate that the LRG distribution does {\it not} show homogeneity
at the claimed HS that was usually determined based on the average
trend of galaxy number count with increasing scale \citep{hogg,scrimgeour,ntelis}.

To quantify the significance of the deviation from homogeneity,
we compare the average $\left<\xi\right>$ and standard deviation
$\sigma_\xi$ estimated from the $\xi$-distribution for the LRG,
random, and mock catalogues.
As shown in Fig.\ \ref{fig:xi_mean_std}, the LRG sample and most of mock catalogues significantly deviate from the range allowed by the random
distribution with spatial homogeneity at $70$--$300~h^{-1}\textrm{Mpc}$ scales ($a$--$b$, $e$--$f$) and the behaviour persists over different redshifts ($b$--$d$, $f$--$h$) for both averages and dispersions.
At $z_c=0.35$ and $R=300~h^{-1}\textrm{Mpc}$ ($f$), the deviations are $11\sigma$ and $17\sigma$ for $\left<\xi\right>$ and $\sigma_\xi$,
respectively; this means statistically {\it impossible} to be realized in the random distribution. Note that the deviation becomes larger at smaller scales.

\begin{figure}
\includegraphics[width=83mm]{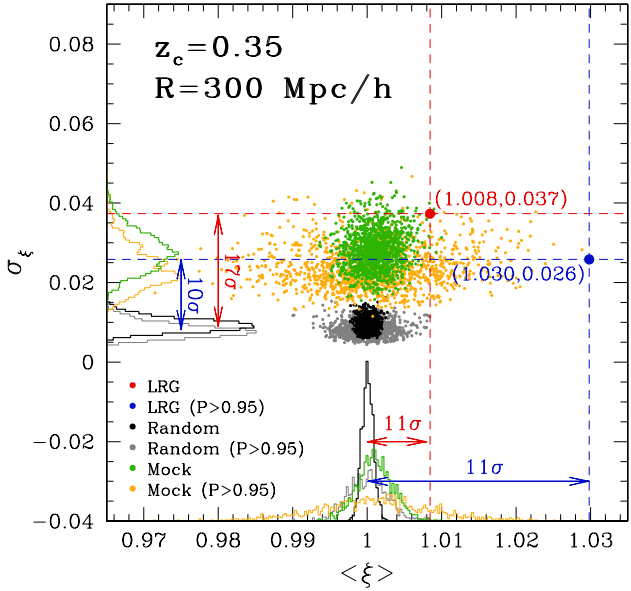}
\caption{Plot of $\left<\xi\right>$ versus $\sigma_\xi$ for the $z_c=0.35$ slice at
         $R=300~h^{-1}\textrm{Mpc}$, which is the same as Fig.\ \ref{fig:xi_mean_std}$f$.
         Here results obtained with galaxies or data points with volume-completeness $P>0.95$ have been added
         for LRG (blue dot with dashed lines and numbers), 1,000 random (grey) and 1,296 mock catalogues
         (green dots). The corresponding histograms for the random and mock catalogues are shown
         on the axes with the same code.
         }
\label{fig:xi-mean_std_P95}
\end{figure}

The mock catalogues also show larger dispersions in both $\left<\xi\right>$ and
$\sigma_\xi$ than the random ones, which is another evidence for the fact
that the $N$-body clustering is no longer homogeneous even at $300~h^{-1}\textrm{Mpc}$ scales.
The LRG results also seem to somewhat deviate from the mock results
at large scales ($3.0\sigma$ for $\left<\xi\right>$ and
$1.9\sigma$ for $\sigma_\xi$; Fig.\ \ref{fig:xi_mean_std}$f$).
Considering the dense overlaps of survey areas of mock catalogues, however,
we can expect that the scatters in $\xi$ statistics become larger for
independent mock samples. In this sense, the LRG deviations are allowed
to occur statistically in the $N$-body simulation\footnote{The $\left<\xi\right>$-distribution for mock catalogues implies that the individual $\left<\xi\right>$ values fluctuate within the range of $0.99$--$1.01$ (Fig.\ \ref{fig:xi_mean_std}$f$). If the mock catalogues mimic the LRG sample well, we can expect that the $\xi$-averages (at $R=300~h^{-1}\textrm{Mpc}$ scale) estimated from many independent LRG-like samples will also fluctuate around such a range. Therefore, the large deviation of $\left<\xi\right>$ from the unit value for the LRG sample can be considered as a natural phenomenon that can happen statistically.}.
Figure \ref{fig:xi_mean_std}$b$--$d$ shows that in the $\left<\xi\right>$--$\sigma_\xi$ plane the three independent LRG data
can occur in arbitrary location of the region occupied by the mock data;
the similar locations of the LRG data for $R=300~h^{-1}\textrm{Mpc}$
in Fig. \ref{fig:xi_mean_std}$f$--$h$ could be due to severe overlapping
of the LRG data in the redshift direction for such a huge radius.

Therefore, we {\it conclude} that the LRG distribution is consistent with the current paradigm of cosmology.
The large-scale inhomogeneity of LRG target selection
may generate spurious inhomogeneity in the LRG distribution.
Since all the LRG results are consistent with the mock results,
we expect that the imperfect target selection does not
affect our main conclusion (see \citealt{hogg}).

In the previous analysis, we included galaxies that are located around the edge of the survey region, and thus some $\xi$ measurements may be less reliable compared with those with higher volume-completeness. In Fig.\ \ref{fig:xi-mean_std_P95},
we show plots in the $\left<\xi\right>-\sigma_\xi$ plane for $z_c=0.35$ and $R=300~h^{-1}\textrm{Mpc}$ obtained with galaxies (or data points) with the volume-completeness ($P$) larger than $0.95$ (excluding galaxies near the survey edge), together with the previous results presented in Fig. \ref{fig:xi_mean_std}$f$.
Due to the smaller number of data points with $P>0.95$, the $\left<\xi\right>$-distributions for random and mock catalogues show larger scatters (grey and yellow dots and histograms). The average value of $\xi$ for LRG with $P>0.95$ significantly deviates from the homogeneity expectation
by $11\sigma$. On the other hand, the $\sigma_\xi$-distributions for random and mock catalogues have shifted to the smaller value, and the position of the LRG $\sigma_\xi$ deviates by $10\sigma$, which is smaller than the previous value obtained by 
including data points around the survey edge because more reliable data points were used in this statistics. Comparison of the two cases of including and excluding the data points around the survey edge suggests that the general behaviour of galaxy number count statistics is maintained in both cases, which makes our conclusion robust.

%
%
\section{Conclusion}

Our study shows that the large-scale structure revealed in the SDSS LRG sample is {\it not} spatially
homogeneous even at $300~h^{-1}\textrm{Mpc}$ scale in radius, substantially deviating from the
expected distribution with homogeneity, see Figs.\ \ref{fig:xi_mean_std} and \ref{fig:xi-mean_std_P95}.
For LRG data there is no HS found yet.
Therefore, the claimed HS at $70~h^{-1}\textrm{Mpc}$ based on similar data \citep{hogg} is disputed.
To determine the value of HS (if it exists), it is essential
to analyse data with larger survey volume; $300~h^{-1}\textrm{Mpc}$ is near maximal radius scale available in the LRG data especially in the redshift direction.
We defer the analysis with more recent data (e.g., SDSS-III BOSS survey; \citealt{alam}) for future work.

If the CP is not in the sky, where is it then in the Universe? The current concordance cosmology theoretically based on the CP in the {\it early} era is generally accepted to be quite successful in explaining most of the cosmological observations. Our homogeneity test shows that the theoretical model prediction is also {\it consistent} with the observation, thus is {\it not} homogeneous compared with the random one. If that is the case the CP may stay in the theoretical foundation of the modern physical cosmology (in the early era), but not in the sky (i.e., not in the present epoch, nor in the observed lightcone). That is, the celebrated modern cosmology paradigm does {\it not} demand the actual presence of CP in our observed sky. This conclusion opens a new possibility in theoretical cosmology demanding careful study of light propagation in nonlinear clustering stage of the world model \citep{Ellis-2008}.

\section*{Acknowledgements}
\addcontentsline{toc}{section}{Acknowledgements}

C.G.P. was supported by Basic Science Research Program through the National
Research Foundation of Korea (NRF) funded by the Ministry of Science, ICT
and Future Planning (No.\ 2013R1A1A1011107).
H.N. was supported by NRF funded by Ministry of Science, ICT and Future Planning (No. 2015R1A2A2A01002791).
J.H. was supported by Basic Science Research Program through the NRF of Korea funded by the Ministry of Science, ICT and Future Planning (No.\ 2016R1A2B4007964).








\appendix
\section*{Appendix}
\label{sec:appendix}
\setcounter{equation}{1}    

{\it The SDSS LRG sample.}
We use the non-official SDSS LRG sample \citep{kazin} which can be obtained
from the internet link, http://cosmo.nyu.edu/$\sim$eak306/SDSS-LRG.html.
The sample is composed of luminous galaxies selected from the SDSS
seventh and final data release (DR7) \citep{sdss-lrg}, and is found to be in agreement
with the official LRG sample \citep{kazin}.

Here, the full DR7 LRG sample with redshift $0.16 < z < 0.47$
and $g$-band absolute magnitude range of $-23.2 < M_g < -21.2$ has been used.
In Fig.\ \ref{fig:lrg}$b$, we have applied the Hammer-Aitoff equal-area
projection \citep{hammer} to plot 26,285 galaxies on the sky.
Given right ascension ($\alpha$) and declination ($\delta$)
in equatorial coordinates, the projection onto $(x,y)$-plane is defined as
\begin{equation}
   x=2\gamma\cos\left(\frac{\alpha}{2}\right)\cos\delta, \quad
   y=\gamma\sin\delta
\end{equation}
where
\begin{equation}
\gamma=\left(\frac{2}{1+\sin\left(\frac{\alpha}{2}\right)
                \cos\delta}\right)^{1/2}.
\end{equation}
As well as the angular position and redshift,
the LRG sample also contains the additional information for each galaxy
such as the sector completeness ($w_s$) and the fibre collision weight ($w_f$).
For usual galaxies, the unit value of the fibre collision weight is assigned ($w_f=1$).
Sometimes target galaxies are separated by less than $55^{\prime\prime}$, the angular diameter covered by the spectroscopic fibre,
and they cannot be observed simultaneously due to fibre collision.
In such cases, only one galaxy has the priority for spectroscopic observation and the fibre collision weight larger than unity
is assigned ($w_f > 1$) \citep{edr}.

{\it Reconstructing angular selection function.}
The observation of SDSS galaxy redshift survey has not been performed
uniformly within the survey region due to several reasons.
The survey region is composed of overlapping tiles
and masked area due to bright stars and objects, fibre priority, bad
fields, and so on. A unique set of tiles covering any area of sky is called
a sector, and each sector is assigned the sector completeness determined by the fraction of galaxies with the successful spectroscopic observation relative to the total objects within the sector.

\setcounter{figure}{5}    
\begin{figure}
\includegraphics[width=85mm]{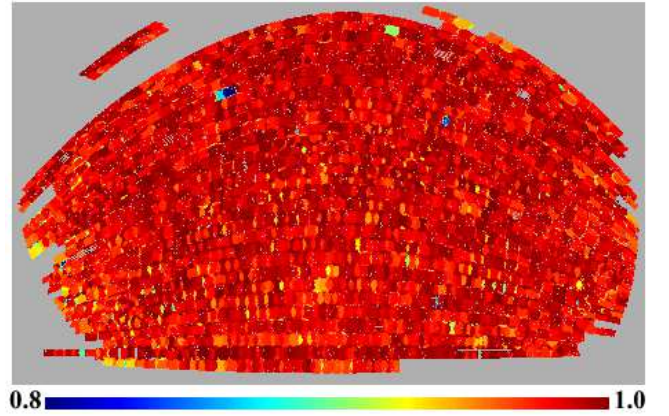}
\caption{Angular selection function of the SDSS DR7 LRG sample.
         The colour indicates the sector completeness ($w_s$). }
\label{fig:angsel}
\end{figure}

Here, the survey completeness in angular direction (angular selection function)
is reconstructed using the SDSS DR7 survey boundary map with mask information provided in \cite{choi} and
the sector completeness information contained in the random point
distribution with 1.7 million data points in \cite{kazin}
(Fig.\ \ref{fig:angsel}).
In representing the angular selection function on the sky,
we use the HEALpix software \citep{healpix} which provides the equal-area
pixelisation on the sky with the total number of pixels
$N_\textrm{pix}=12 N_\textrm{side}^2$,
where $N_\textrm{side}$ is the resolution parameter.
$N_\textrm{side}=2048$ is sufficient for our purpose of mapping
the angular selection function.

{\it Generating random and mock catalogues.}
In order to establish the criterion for the spatial homogeneity,
it is essential to generate the random catalogues, each containing the Poisson-distributed
data points. A random catalogue is generated as follows.
First, a redshift $z$ is randomly drawn in the range $0.16 \le z \le 0.47$
with a probability function shown in Fig.\ \ref{fig:lrg}$a$.
Second, angular position $(\alpha,\delta)$ on the sky is randomly drawn
using the angular selection function map as the probability function.
Each random data point is assigned the sector completeness ($w_s$).
We assign the uniform fibre collision weights to all the random points ($w_f=1$).
In this way, we generate a random point catalogue which includes the same
number of galaxies as in the LRG sample, and 1,000 random catalogues
in total.

The mock catalogues mimicking the LRG sample are generated in a similar way.
\cite{kim} performed the Horizon Run 3 $N$-body simulation using $374$ billion particles in a volume of $(10.815~h^{-1}\textrm{Gpc})^3$, which allows to resolve galaxy-size halos with mean particle separation of $1.5~h^{-1}\textrm{Mpc}$.
A set of 27 all-sky mock surveys (designed for SDSS-III) along the past
lightcone out to $z=0.7$ is publicly available (http://sdss.kias.re.kr/astro/Horizon-Runs/Horizon-Run23.php).
The cosmological model used in the simulation
is the $\Lambda$ cold dark matter ($\Lambda\textrm{CDM}$) dominated universe with $\Omega_M=0.26$, $\Omega_B=0.044$,
$\Omega_\Lambda=0.74$, $n_s=0.96$, $h=0.72$, and $\sigma_8=0.79$,
where $\Omega_M$, $\Omega_B$, $\Omega_\Lambda$ are the current matter, baryon, dark energy density parameters, respectively, $n_s$ the spectral index of primordial scalar-type perturbation, $\sigma_8$ the amplitude of
the matter fluctuations at $8~h^{-1}\textrm{Mpc}$ scale.
The angular positions of the survey centre are chosen based on the
HEALpix pixelisation of $N_\textrm{side}=2$.
Thus, for each halo catalogue, 48 LRG mock catalogues are generated with
individual survey area uniformly separated (by about $30^\circ$) but significantly overlapped on the sky, and 1296 ($=27\times 48$) catalogues are made in total.
We make each mock catalogue by considering the LRG angular selection function and the redshift distribution. For each halo within the LRG survey area, we draw a random number from $[0,1]$ with uniform distribution. If the random number is smaller than the value of angular selection function at the halo's angular position, the halo is chosen, otherwise it is discarded. From all the chosen halos within the survey area, massive halos with the same number of LRG within a given redshift interval have been extracted in decreasing order of halo mass. A mock catalogue is obtained by repeating the process over all the redshift intervals.
In the LRG mock survey, we also assign the sector completeness ($w_s$) to each halo and the uniform fibre collision weights
to all the data points ($w_f=1$).

{\it Counting galaxies within a sphere.}
We assume $\Lambda\textrm{CDM}$ dominated universe to calculate
the comoving distance to a galaxy at redshift $z$
\begin{equation}
   r(z)= \frac{c}{H_0}
   \int_0^z \frac{dz}{\sqrt{\Omega_M (1+z)^3 +\Omega_\Lambda }},
\end{equation}
where $c$ is the speed of light and $H_0$ is Hubble's constant.
We assume $\Omega_{M}=0.27$ and $\Omega_\Lambda=0.73$.
Given a sphere of comoving radius $R$ at redshift $z$ or at distance $r_0=r(z)$,
the angular radius of the sphere on the sky is given by
$\theta_R=\sin^{-1} (R/r_0)$.
We count galaxies within the sphere from the LRG sample and compare the number count with that expected from the homogeneous distribution.
Each galaxy contributes $w_f / w_s$ to the galaxy number counts.
The random point distribution with 100 times larger number of points than the LRG catalog
is used to estimate the expected number of galaxies within a sphere.
Each random point contributes $0.01 w_f/w_s$
to the count. The scaled $\mathcal{N}(R)$ is obtained from the sum of all the LRG contributions
divided by that of random point distribution within the sphere of radius $R$.

\begin{figure}
\centering
\includegraphics[width=80mm]{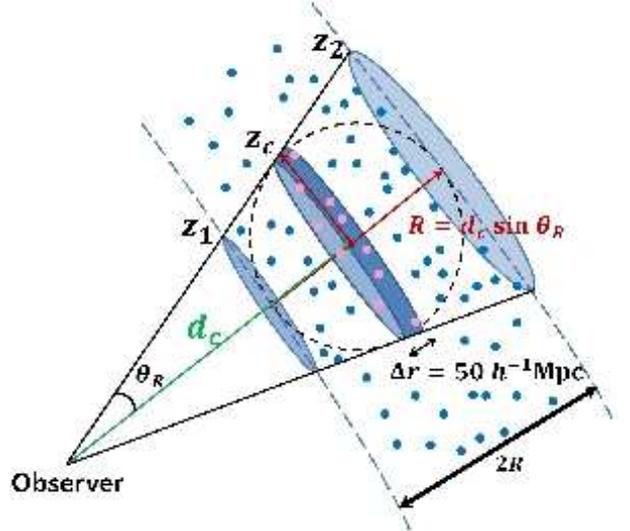}
\caption{Geometry of a truncated cone.
         The picture describes a thin slice with central redshift $z_c$
         (at comoving distance $d_c$)
         and thickness $\Delta r=50~h^{-1}\textrm{Mpc}$ together with galaxies
         inside the slice (pink dots), a sphere of radius $R$ whose centre
         is located at the slice centre but with a galaxy's angular position,
         and a truncated cone that is circumscribed about the sphere as seen
         by an observer, with angular radius $\theta_R$.
         The upper and base sides of the truncated cone correspond to
         redshifts $z_1$ and $z_2$, respectively.
         The blue dots indicate galaxies within a slice with $2R$ thickness,
         the height of the truncated cone in comoving distance.
        }
\label{fig:geom_trc}
\end{figure}

\begin{figure*}
\centering
\includegraphics[width=85mm]{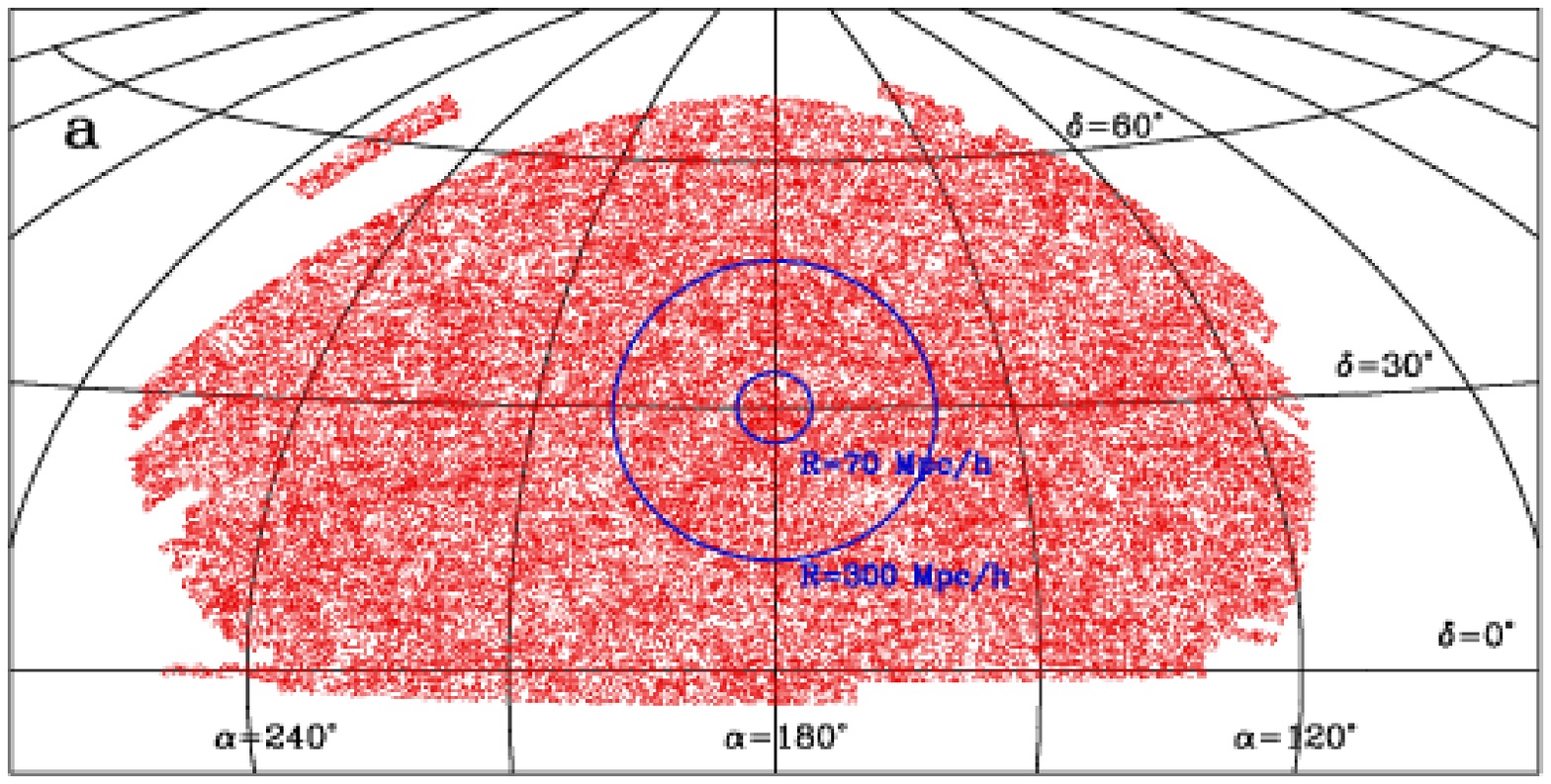}
\includegraphics[width=85mm]{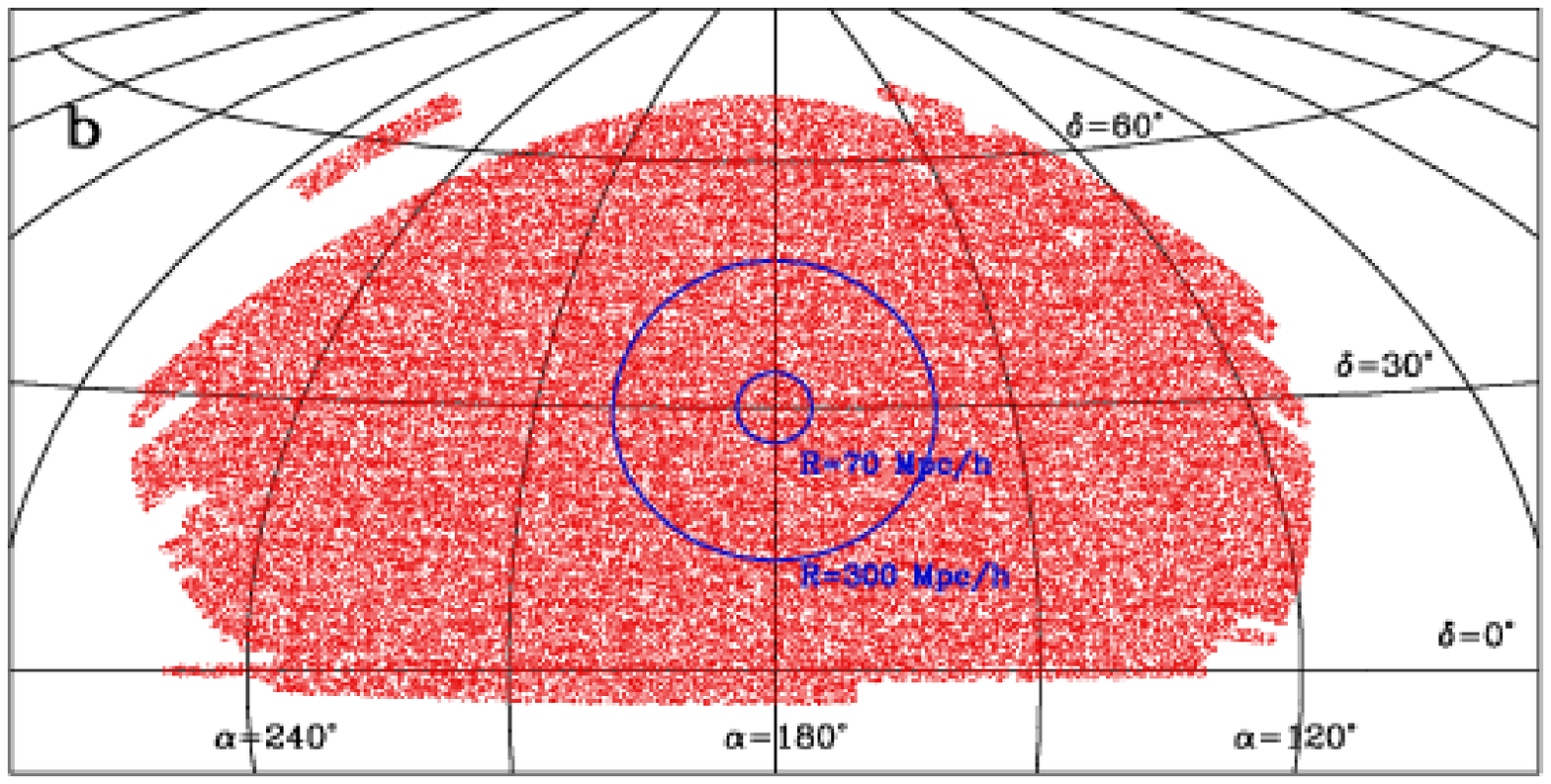}
\includegraphics[width=85mm]{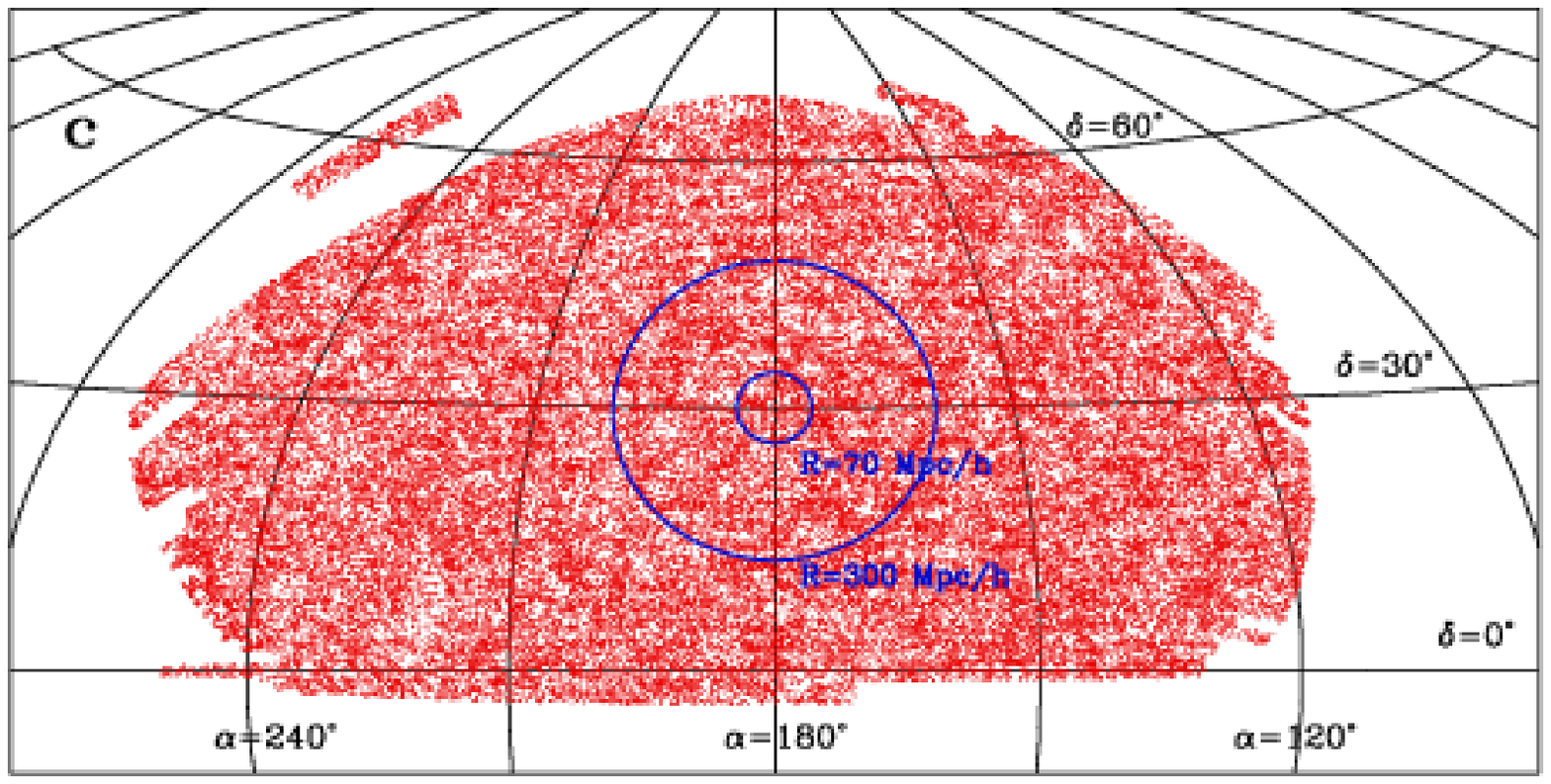}
\includegraphics[width=85mm]{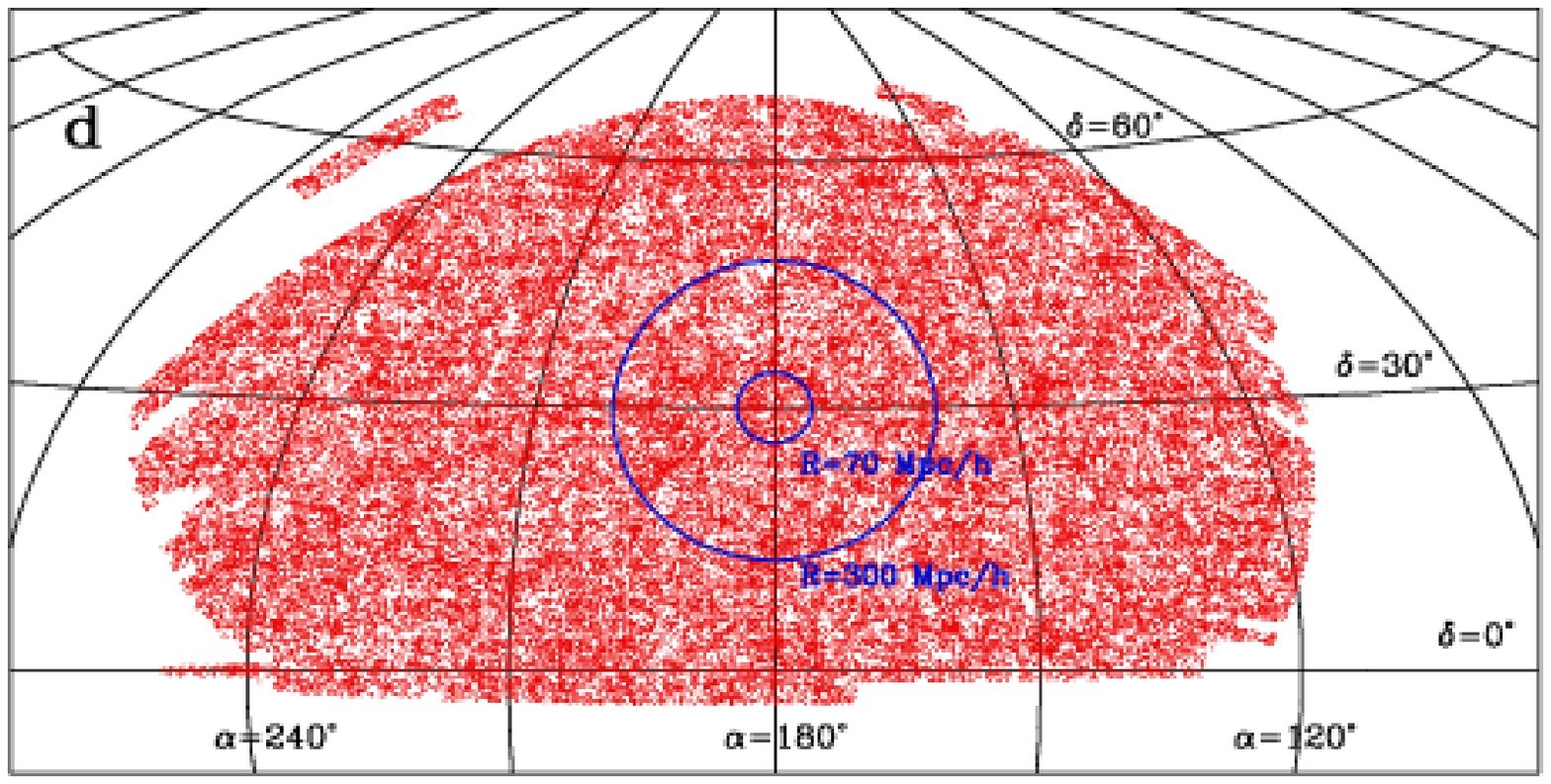}
\caption{Angular distributions of data points
         at $0.235 < z < 0.470$ for ($a$) the SDSS LRG,
         ($b$) one random, and mock catalogues with
         ($c$) the maximum and ($d$) minimum dispersions.
         Each distribution corresponds to the slice defined by
         the central redshift $z_c=0.35$ and the thickness of
         $2R=600~h^{-1}\textrm{Mpc}$. The slice has been a bit cut off
         by the survey boundary at the maximum distance ($z=0.47$).
         Corresponding maps of $\xi$ are presented in Fig.\ \ref{fig:xi-maps}.
         }
\label{fig:lrg_R300}
\end{figure*}

Generally, the volume of a sphere centred at each galaxy is incomplete due to the survey boundary and masked area. The volume-completeness for each measured $\mathcal{N}$ is obtained by comparing
the volume of the sphere contained within the survey region with that of a complete sphere, centred at the location of a galaxy. The volume of a sphere is estimated by integrating the volume elements
over the direction of HEALpix pixels with $N_\textrm{side}=2048$
penetrating the sphere.
The distances to the near and far ends of the penetrating line are
\begin{equation}
   r_{\pm}=r_0 \left[\cos\theta \pm \sqrt{\sin^2 \theta_R - \sin^2 \theta} \right]
\end{equation}
where $\theta$ is the angular separation between the centre of a sphere
and a line-of-sight direction penetrating the sphere.
The volume element for each pixel is given by
\begin{equation}
   v_\textrm{pix}=\frac{\Omega_\textrm{pix}}{3}(r_{+}^3-r_{-}^3)
\label{eq:vpix}
\end{equation}
where $\Omega_\textrm{pix}=4\pi/N_\textrm{pix}$.
The total volume of the sphere is the sum of all the volume elements
within the survey region.
Sometimes, the sphere is cut off by the survey boundary at the minimum/maximum distance ($r_\textrm{min}$/$r_\textrm{max}$) from us. In that cases, we set $r_{-}=r_\textrm{min}$ or $r_{+}=r_\textrm{max}$.

{\it Counting galaxies within a truncated cone.}
In the $\xi$ measurement, the count-in-redshift-range method does not
need to use the random point distribution to correct the bias in radial direction.
The method estimates only the number density of galaxies within the truncated cone and within
the whole slice which is determined by the redshift range ($z_1 < z < z_2$) set by the upper
and base sides of the truncated cone (Fig.\ \ref{fig:geom_trc}).
As in the count-in-sphere method, we consider the sector completeness and the fibre collision weight in counting the galaxies.
We estimate the volume of the truncated cone contained
within the survey region using the method in the case of count-in-sphere.
That is, we compute the partial volume of the truncated cone by adding up individual volume elements $v_\textrm{pix}$ [Eq.\ (\ref{eq:vpix})] over the pixelised angular area on the sky enclosed by the partial truncated cone, where we choose the same resolution parameter $N_\textrm{side}=2048$ for the HEALpix pixelisation as in the case of count-in-sphere.
However, in this case the $r_{-}$ and $r_{+}$ are comoving distances to
$z_1$ and $z_2$, respectively, for a given central redshift $z_c$ and the scale radius $R$.
To estimate the volume-completeness, we also need the volume of the complete truncated cone, which is given by
\begin{equation}
   V_\textrm{trc}=\frac{4\pi}{3} \left(r_{+}^3 -r_{-}^3 \right)
                   \sin^2 \left( \frac{\theta_R}{2} \right).
\end{equation}

{\it Generating $\xi$-maps.}
The measurement of $\xi$ has been done at the slice centre and angular
position of each data point in the LRG, random, and mock samples.
In the $\xi$-maps shown in Fig.\ \ref{fig:xi-maps}, however,
we calculate $\xi$ at angular positions centred on HEALPix pixels
(with $N_\textrm{side}=256$) within the survey region in order to obtain
the continuous maps. Other details are exactly the same as in the method
described in the text.
Figure \ref{fig:lrg_R300} shows angular distributions of data points
that were used in generating Fig.\ \ref{fig:xi-maps}.


\bsp	
\label{lastpage}
\end{document}